\documentclass[aps,prd,superscriptaddress,nofootinbib,11pt]{revtex4}
\usepackage[english]{babel}
\usepackage[utf8]{inputenc}
\usepackage{graphicx}   
\usepackage{slashed}
\usepackage{epstopdf}
\usepackage{verbatim}   
\usepackage{color}      
\usepackage{subfigure}  
\usepackage{multirow}
\usepackage{hyperref}   
\usepackage{float}
\usepackage{epsfig,rotating}
\usepackage{amsmath,amssymb}
\usepackage{dsfont}
\restylefloat{table}
\numberwithin{equation}{section}

\newcommand{\vx}{\vec{x}}

\newcommand{\vk}{\vec{k}}

\newcommand{\be}{\begin{equation}}
\newcommand{\ee}{\end{equation}}
\newcommand{\bea}{\begin{eqnarray}}
\newcommand{\eea}{\end{eqnarray}}

\newcommand{\ket}[1]{|#1\rangle}
\newcommand{\bra}[1]{\langle#1|}

\newcommand{\X}{\mathcal X}
\newcommand{\M}{\mathcal M}

\begin{document}
\title{Condensate decay in a radiation dominated cosmology.}

\author{Shuyang Cao}
\email{shuyang.cao@pitt.edu} \affiliation{Department of Physics, University of Pittsburgh, Pittsburgh, PA 15260}
\author{Daniel Boyanovsky}
\email{boyan@pitt.edu} \affiliation{Department of Physics, University of Pittsburgh, Pittsburgh, PA 15260}

 \date{\today}

\begin{abstract}
We study the decay of a homogeneous condensate of a massive scalar field of mass  \emph{m}  into massless fields  in thermal equilibrium  in a radiation dominated cosmology. The model is a \emph{proxy} for the non-equilibrium dynamics of a misaligned axion condensate decaying into radiation.  After consistent field quantization in the cosmological background, we   obtain the linearized causal equations of motion for a homogeneous condensate including the finite temperature self-energy corrections up to one loop. The  dynamical renormalization group is implemented  to obtain the time dependent relaxation rate that describes the decay dynamics   of the condensate amplitude from stimulated emission and recombination of massless quanta in the medium. It is explicitly shown  that a simple friction term in the equation of motion  does not  describe correctly  the decay of the condensate. During the super-Hubble regime,    relevant for ultralight dark matter,  the condensate amplitude decays as $e^{-\frac{g^2}{10} t^2\,\ln(1/mt)}$. In the sub-Hubble regime the amplitude decays as $e^{-\gamma(t;T(t))/2}$ with $T(t)=T_0/a(t)$, therefore the finite temperature contribution to the decay rate vanishes fast during the expansion.    A main conclusion is that  a phenomenological friction term is inadequate to describe the decay in the super-Hubble regime,  the decay function $\gamma(t)$  is always \emph{smaller} than that from a local friction term as a consequence of the cosmological expansion. For ultra light dark matter, the time scale during which transient dynamics is sensitive to cosmological expansion and  a local friction term is inadequate, is much longer. A friction term always     \emph{underestimates} the time scale of decay in the sub-Hubble case.
\end{abstract}


\maketitle

\section{Introduction}

The decay of the expectation value of a scalar (or pseudoscalar) field, namely a condensate, plays a fundamental role in inflationary and post-inflationary cosmology\cite{kolb}. Two important examples are the inflaton, a scalar field whose homogeneous condensate drives inflation, and   axionic dark matter.   Axions are pseudoscalar particles originally introduced as possible solutions to the strong CP problem in QCD\cite{PQ,weinaxion,wil},  may be  produced non-thermally in the Early Universe, for example by a misalignment mechanism, and are    very compelling  cold dark matter candidates\cite{pres,abbott,dine}. Axion-like particles  may also be accomodated  in extensions beyond the standard model\cite{banks,ringwald,marsh,sikivie1,sikivie2},  in particular as candidates for ultra light dark matter\cite{fuzzy,uldm}. The relaxation of such a condensate as a consequence of its decay into other particles is typically described phenomenologically via a local friction term\cite{kolb,berera}. If these scalar (or pseudoscalar) condensates account for the dark matter component,   understanding in depth their  relaxation   in an expanding cosmology is  of fundamental importance in the determination of the dark matter abundance.

The decay of a scalar condensate is also relevant within the context of reheating post-inflation\cite{rehe1}-\cite{rehe6}, where the decay of the inflaton condensate into other degrees of freedom results in a large population of the latter that eventually thermalizes via collisional (or other) processes giving rise to  the post-inflationary radiation dominated era. Shortcomings in the description of the relaxation of the amplitude of the scalar condensate in terms of a local friction term, even in Minkowski space time, have been discussed in refs.\cite{fric1,fric2}.

Cosmological expansion with the concomitant lack of a time-like Killing vector allows processes that are kinematically forbidden by energy-momentum conservation in Minkowski space-time. Previous study of inflaton decay in a de Sitter cosmology\cite{inflaton} revealed several important   processes that are unavailable in Minkowski space-time. Single particle decay in a radiation dominated cosmology revealed\cite{herring} novel contributions as a consequence of the cosmological expansion, without counterparts in Minkowski space time.  These lead to decay laws which are very different from simple exponentials in time as is the usual result in Minkowski space-time from S-matrix theory which implements strict energy-momentum conservation in the infinite time limit. Along with the lack of a time-like Killing vector in an expanding cosmology, the finite Hubble time scale leads to a time-energy uncertainty which relaxes the kinematic constraint of strict energy conservation, thereby allowing novel decay and scattering  channels unavailable in Minkowski space time\cite{boyrai}. Some of  these novel features were also recognized in  refs.\cite{audi,vilja}.

 The dynamics of axion condensates decaying into   photons in \emph{Minkowski space-time} was recently studied in ref.\cite{shuboy} where it is explicitly shown how the axion self-energy from photon loops yields a non-local self-energy contribution to the equation of motion of the condensate that determines its decay and thermalization.

 The relaxation  of a condensate as a consequence of decay into or thermalization with other degrees of freedom in a medium is a manifestation of energy transfer between different species. Recent studies\cite{dienes}  revealed  intriguing phenomena as a consequence of  energy transfer between species in an expanding cosmology  that lead to a novel emergent dynamical attractor, a fixed point for the abundances of the various species that arises from a balance between energy transfer amongst them and the cosmological expansion.

   Particle decay is an ubiquitous process. In \emph{Minkowsky space-time} the decay \emph{rate} is obtained in S-matrix theory from the transition probability per unit time,  from an in-state prepared at time $t\rightarrow -\infty$ to an out-state at time $t\rightarrow \infty$. Alternatively it is also obtained from Fermi's Golden rule, where again the infinite time limit is taken, thereby approximating a sinc-type function\footnote{The function $\mathrm{sinc}(x) = \sin(x)/x$.} that reflects the energy-time uncertainty, with an exact delta function yielding energy conservation. In a cosmological background, a \emph{phenomenological} local friction term is often included in the equation of motion of the condensate to describe its   dissipation from  its decay into other degrees of freedom\cite{kolb,berera}. This phenomenological friction term    is typically calculated within the S-matrix framework of Minkowski space-time\cite{berera}. Obviously, taking the infinite time limit, as in the S-matrix calculation of the decay rate in a time dependent, expanding cosmology is at best an approximation that merits deeper scrutiny.

   However, to the best of our knowledge a thorough derivation and solution of the equations of motion for a condensate in an expanding radiation dominated cosmology, including dissipative processes described consistently in terms of the self-energy (namely radiation reaction) has not yet been discussed.

\textbf{Motivation and Objectives:}
Motivated by its importance in the dynamics of  possible dark matter candidates, as well as in the dynamics of post-inflationary reheating and, indirectly  as a mechanism of energy transfer between species,  in this article we study the relaxation of an homogeneous condensate of a scalar (or pseudoscalar) particle via emission and absorption of quanta of a massless field in thermal equilibrium in a radiation dominated cosmology.

This model serves as a \emph{proxy} for the dynamics of a misaligned homogeneous axion condensate decaying into radiation.

 The main objectives of this study are the following:

 \textbf{i:)} to systematically obtain \emph{ab initio} the linearized causal (retarded) equations of motion for the homogeneous expectation value--condensate-- of a scalar (or pseudoscalar) field interacting with massless fields into which it can decay, including self-energy corrections up to one loop in a radiation dominated cosmology.

   \textbf{ii:)} To study the relaxational dynamics of this condensate as a consequence of the emission and absorption of these massless fields which are assumed to be in thermal equilibrium with the radiation background.

   \textbf{iii:)} To explicitly compare the decay dynamics derived from the one-loop self-energy in the radiation dominated cosmological background to the case of Minkowski space time,  and to assess the reliability of a local friction term, obtained from the Minkowski space-time S-matrix,  to describe the dissipative processes from emission and absorption of the massless quanta.

   \vspace{1mm}

   \textbf{Brief summary of results:}

   We quantize the theory in the cosmological background, and obtain the causal equations of motion for an homogeneous condensate by implementing the in-in (Schwinger-Keldysh)\cite{schwinger}-\cite{boylee} formulation of non-equilibrium quantum field theory, including a one-loop self energy correction from the emission and absorption of massless quanta in thermal equilibrium with the background.  The  non-perturbative dynamical renormalization group, originally introduced to obtain amplitude equations in pattern formation\cite{drggold}, and extended to study relaxation phenomena in non-equilibrium quantum field theory\cite{drg1,drg} is implemented to solve the evolution equations and to obtain the decay law of the condensate. We study analytically two different regimes determined by the ratio $m/H(t)=d_H(t)/\lambda_c$ with $m$ the mass of the scalar field of the condensate ($\lambda_c$ the Compton wavelength)  and $H(t)$ the Hubble expansion rate ($d_H$ is the Hubble radius): $m/H \ll 1$ the super-Hubble   and $m/H \gg 1$ the sub-Hubble regimes.
   The super-Hubble regime, more relevant for ultralight dark matter candidates,  is dominated by the zero temperature contribution to the self energy, the amplitude of the condensate decays as  $e^{-\frac{g^2}{10} t^2\,\ln(1/mt)}$ where $g$ is the coupling to massless fields. In this regime a local friction term is completely inadequate to describe the decay of the condensate. In the sub-Hubble regime the condensate amplitude decays as $e^{-\gamma(t;T(t))}$ where $T(t)=T/a(t)$. Although the decay function $\gamma(t,T(t))$ approaches  the \emph{zero temperature} Minkowski space-time decay law $\Gamma t$ with $\Gamma$ the zero temperature S-matrix decay rate in the \emph{asymptotic infinite time limit}, as a consequence of the cosmological expansion it is \emph{always} \emph{smaller} than the Minkowski space time result. As an important consequence, a phenomenological local friction term with  S-matrix decay rate  always \emph{underestimates} the lifetime   of the condensate. Furthermore, our results clearly show that the finite temperature contribution enters the decay law with the redshifted temperature, which obviously invalidates any result that invokes an infinite time limit.

   Taken together, these results indicate that the description of  decay (relaxation) of a homogeneous condensate in terms of a local friction term calculated from the S-matrix in Minkowski space time is in general inadequate.

   A more general lesson drawn from this study is that the formulation to obtain the causal (retarded) equations of motion  introduced here, together with the dynamical renormalization group provide a consistent and systematic framework to obtain and solve the equation of motion  including radiative corrections for scalar or pseudoscalar condensates in cosmology and extract its relaxational dynamics.

  The article is organized as follows: in section (\ref{sec:model}) we introduce the model, in section (\ref{sec:quantization}) we quantize both massive and massless field in the background of a radiation dominated cosmology, analyzing the super and sub-Hubble limits of the mode functions. In section (\ref{sec:timeevol}), we obtain the causal (retarded) equations of motion for an homogeneous condensate implementing the in-in formulation of non-equilibrium quantum field theory including the one-loop self energy contribution from thermal massless fields. In section (\ref{sec:drg}) we implement the dynamical renormalization group to obtain the decay law of the condensate in the super and sub-Hubble limits, analyzing in detail the zero and finite temperature contributions and the renormalization aspects. In section (\ref{sec:discussion}) we discuss caveats in the \emph{proxy} model to axion dynamics,     the scale factor dependence of the interactions,  and possible consequences of slower relaxation on dynamical fixed points in the cosmological evolution of several species. Section (\ref{sec:conclusions}) presents our conclusions and poses further questions. In appendix (\ref{app:eomlap}) we obtain the causal (retarded) equations of motion for the same model in Minkowski space time and solve it via Laplace transform in terms of the spectral representation of the one-loop self-energy, available because of time translational invariance. We obtain the decay law from the complex poles in the (retarded) propagator. In Appendix (\ref{app:drg}) we solve the equations of motion directly in real time by implementing the dynamical renormalization group, thereby showing how this formulation yields the Minkowski space time result. These two appendices allow us to establish a direct comparison between the dynamics during the radiation dominated cosmology and in Minkowski space time, and furthermore, establish  the reliability of the dynamical renormalization group to obtain the
   relaxational dynamics of  cosmological condensates.

\section{The model}\label{sec:model}

We focus on a post-inflationary universe, described by a   spatially flat Friedmann-Robertson-Walker (FRW) cosmology  with the metric in comoving coordinates given by
\be  g_{\mu \nu} = \textrm{diag}(1, -a^2, -a^2, -a^2) \,. \label{frwmetric}\ee The standard cosmology post-inflation is described by  three distinct stages: radiation (RD), matter (MD) and dark energy (DE) domination; we model the latter by a cosmological constant. Friedmann's equation is
\be \Big(\frac{\dot{a}}{a}\Big)^2 = H^2(t) = H^2_0\,\Bigg[ \frac{\Omega_M}{a^3(t)}+ \frac{\Omega_R}{a^4(t)}+ \Omega_\Lambda \Bigg] \,,\label{Hubble} \ee
where the scale factor is normalized to $a_0=a(t_0)=1$ today. We take as representative the following  values of the parameters \cite{wmap,spergel,planck}:
\be H_0 = 1.5\times 10^{-42}\,\mathrm{GeV}~~;~~ \Omega_M = 0.308~~;~~ \Omega_R = 5\times 10^{-5}~~;~~ \Omega_\Lambda = 0.692 \,.\label{cosmopars}\ee
It is convenient to pass from ``comoving time,'' $t$,   to conformal time $\eta$ with $d\eta = dt/a(t)$, in terms of which the metric becomes ($a(\eta)\equiv a(t(\eta))$)
\be  g_{\mu \nu} = a^2(\eta)\, \textrm{diag}(1,-1,-1,-1) \, . \label{conformalmetric} \ee
With (${\,}^{\,'} \equiv \frac{d}{d\eta}$) we find
\be a'(\eta) = H_0\,\sqrt{\Omega_M}\,\Big[r+a+s\,a^4\Big]^{1/2}\,,\label{dera}\ee with
\be r= \frac{\Omega_R}{\Omega_M} \simeq 1.62\,\times 10^{-4}~~;~~ s = \frac{\Omega_\Lambda}{\Omega_M} \simeq 2.25 \,.\label{rands}\ee
Hence the different stages of cosmological evolution, namely radiation domination (RD), matter domination (MD), and dark energy domination (DE), are characterized by
\be a\ll r \Rightarrow \text{RD}~~;~~ r \ll a \lesssim 0.76 \Rightarrow \text{MD} ~~;~~ a > 0.76 \Rightarrow \text{DE} \,. \label{stages}\ee
In the standard cosmological picture and the majority of the well motivated variants, most of the interesting particle physics processes occur during the RD era, therefore, we focus our attention on this epoch, during which the scale factor can be written as
\be a(\eta)= H_R \,\eta   ~~;~~ H_R= H_0\,\sqrt{\Omega_R}\,.  \label{aofetarm}\ee During the (RD) stage, the Hubble rate of expansion in conformal time is
\be H(\eta) = \frac{a'(\eta)}{a^2(\eta)} = \frac{1}{H_R\,\eta^2} \,,\label{hubble} \ee
and   the relation between conformal and comoving time is given by
 \be \eta = \Big( \frac{2\,t}{H_R}\Big)^{\frac{1}{2}} \Rightarrow a(t) = \Big[ 2\,t H_R\Big]^{\frac{1}{2}}\,. \label{etaoft} \ee

We consider  two interacting scalar fields $\phi_1,\phi_2$ in  the FRW cosmology determined by the metric (\ref{frwmetric}),  with action given by\cite{birrell,parker}
\be  S   =    \int d^4 x \sqrt{|g|} \Bigg\{\frac{1}{2} g^{\mu\nu}\,\partial_\mu \phi_1 \partial_\nu \phi_1-\frac{1}{2} \big[m^2_1 -\xi_1\,R\big]\phi^2_1 + \frac{1}{2} g^{\mu\nu}\,\partial_\mu \phi_2 \partial_\nu \phi_2-\frac{1}{2} \big[m^2_2 -\xi_2\,R\big]\phi^2_2   -   \lambda \phi_1 :\phi^2_2:  \Bigg\}
\label{action}\ee where
\be R= 6\Big[\frac{\ddot{a}}{a}+\Big(\frac{\dot{a}}{a} \Big)^2 \Big] \label{ricci}\ee is the Ricci scalar,  and $\xi_{1,2}$ are couplings to gravity, with $\xi=0, 1/6$ corresponding to minimal or conformal coupling, respectively. We identify $\phi_1$ with  the field associated with the  system, for example the axion, and $\phi_2$ as the ``bath'' which will be traced over to yield
the non-equilibrium effective action for $\phi_1$ from which we will extract the equations of motion.

This model is intended as a   \emph{proxy} to that of an axion-like particle that couples to photons via a topological Chern-Simons term, a consequence of the $U(1)$ anomaly. The  model described by (\ref{action}) provides a simpler arena to introduce the methods for quantization in the cosmological background, obtaining the causal equations of motion and their solution via the dynamical renormalization group bypassing the complications  associated with  the topological nature of the Chern-Simons coupling to gauge fields in the cosmological background. While the model (\ref{action}) describes the decay into massless fields, a feature of axion-like particles decaying into photons, there are important caveats associated with the latter case that will be discussed in detail in   section (\ref{sec:discussion}).

The normal ordering symbol in the Lagrangian density (\ref{action}) stands for the definition
\be :\phi^2_2: =  \phi^2_2 - \langle  \phi^2_2 \rangle \,,\label{norord}\ee where the expectation value is in the equilibrium density matrix describing the massless scalar field $\phi_2$ in thermal equilibrium with the radiation background.

Expressing the action of Eq.~(\ref{action}) in terms of the comoving spatial coordinates and conformal time,   rescaling the fields as\cite{birrell,parker}
\be \phi_{1,2}(\vec{x},t) = \frac{\chi_{1,2}(\vec{x},\eta)}{a(\eta)} \,, \label{rescale}\ee  with
\be R= 6 \, \frac{a''(\eta)}{a^3(\eta)}\,, \label{Ricciconformal} \ee where the primes now refer to derivatives with respect to conformal time, and neglecting surface terms that do not modify the equations of motion,
the action becomes
\be S = \int d^3x \, d\eta  \biggl\{\sum_{j=1,2}\Bigl[\tfrac{1}{2}  \,\Big(\frac{\partial \chi_j}{\partial \eta} \Big)^2 -\tfrac{1}{2} \,\bigl( \nabla \chi_j \bigr)^2 -\tfrac{1}{2} \chi^2_j\, \mathcal{M}^2_j(\eta) \Bigr] -   \lambda \,a(\eta)\,\chi_1 \,:\chi^2_2:\,  \biggr\}\,
 \label{conformalaction}\ee
  with the effective, time dependent masses given by \be
  \mathcal{M}^2_j(\eta) =m^2_j \,a^2(\eta)- \frac{a''(\eta)}{a(\eta)}\,(1-6\xi_j)  \,.  \label{massconfor} \ee

In a radiation dominated FRW cosmology with $a(\eta)$ given by (\ref{aofetarm}), the Ricci scalar $R=0$,  and
\be \mathcal{M}^2_{j}(\eta) = m^2_j\,a^2(\eta) ~~;~~j=1,2 \,. \label{massconformal}\ee We study the case of massless bath fields, namely  $m_2=0$, hence massless bosonic fields in a radiation dominated FRW cosmology are equivalent to conformally coupled massless fields, a situation which bears relevance in the case of gauge fields.

\vspace{1mm}

\section{Quantization:}\label{sec:quantization}
The conformal time action (\ref{conformalaction}) yields the canonical momenta
\be \pi_j(\vx,\eta) = \frac{d\chi_j(\vx,\eta)}{d\eta} \,,\label{canonical}\ee and quantization is achieved by the canonical commutation relation of the respective operators, namely
\be \Big[\pi_i(\vx,\eta),\chi_j(\vx^{\,\,'},\eta)\Big] = -i\delta_{ij}\,\delta^3 (\vx-{\vx}^{\,'}) \,. \label{commrel}\ee

\textbf{Massless fields:}
  For the  massless, real scalar field $\phi_2$ (namely the conformally rescaled $\chi_2$), the free field equations of motion in conformal time and comoving coordinates obtained from the action (\ref{conformalaction}) are
\be \frac{\partial^2 \chi_2(\vx,\eta)}{\partial \eta^2} - \nabla^2 \chi_2(\vx,\eta) = 0 \label{masslesEOM} \ee which are the same as for a massless field in Minkowski space time but with time replaced by conformal time $\eta$. Therefore, the Heisenberg free field operator  is given by
\be \chi_2(\vx,\eta) = \frac{1}{\sqrt{V}}\sum_{\vk} \frac{1}{\sqrt{2k}}\Big[ b_{\vk} \, e^{-i k\eta}\,e^{i \vk \cdot \vx} + b^\dagger_{\vk} \, e^{i k\eta}\,e^{-i \vk \cdot \vx}\Big] \label{chi2exp} \ee where $V$ is the comoving quantization volume, $\vk;\vx$ are  comoving momenta and coordinates, and $b_{\vk},b^\dagger_{\vk}$ are time independent annihilation and creation operators with canonical commutation relations.

\vspace{1mm}
\textbf{Massive fields:} For the massive $\phi_1$ ($\chi_1$) field, the free-field equation of motion from the action (\ref{action})  is
\be \frac{\partial^2 \chi_1(\vx,\eta)}{\partial \eta^2} - \nabla^2 \chi_1(\vx,\eta) + \mathcal{M}^2_1(\eta)\,\chi_1 (\vx,\eta) = 0\,.  \label{massiveEOM} \ee We expand the field in Fourier modes in comoving coordinates, writing
\be \chi_1(\vx,\eta) =  \frac{1}{\sqrt{V}}\sum_{\vk}  \Big[ a_{\vk} \,\mathcal{U}_{k}(\eta)\,e^{i \vk \cdot \vx} + a^\dagger_{\vk} \, \mathcal{U}^*_k(\eta)\,e^{i \vk \cdot \vx}\Big] \,, \label{chi1exp} \ee where the mode functions $\mathcal{U}_{k}(\eta)$ are solutions of the following equations of motion
\be \frac{d^2\,\mathcal{U}_k(\eta)}{d\eta^2} + \Big(k^2+ m^2\,H^2_R \,\eta^2\big)\,\mathcal{U}_k(\eta) =0 \,,\label{Umodes}\ee with Wronskian condition\footnote{We denote the Wronskian with respect to a variable $x$ as $\mathcal{W}_x[f_1(x),f_2(x)]\equiv \frac{d f_1(x)}{dx}\,f_2(x)-\frac{d f_2(x)}{dx}\,f_1(x)$.}
\be \mathcal{W}_{\eta}\Big[\mathcal{U}_k(\eta),\mathcal{U}^*_k(\eta)\Big] \equiv \frac{d \,\mathcal{U}_k(\eta)}{d\eta}\,\mathcal{U}^*_k(\eta) - \frac{d \,\mathcal{U}^*_k(\eta)}{d\eta}\,\mathcal{U}_k(\eta) = -i \,,\label{wronskian} \ee which guarantees that the time independent  annihilation and creation operators $a_{\vk}, a^\dagger_{\vk}$ obey canonical commutation relations.
Introducing the dimensionless variables
\be x= \sqrt{2 m H_R}\,\,\eta ~~;~~ \kappa^2 = \frac{k^2}{2mH_R}\,,\label{xvar}\ee  the equation of motion (\ref{massiveEOM}) becomes
\be \Big[\frac{d^2}{dx^2} + \Big(\kappa^2+ \frac{x^2}{4}\Big)\Big]\,U_\kappa(x) =0 \,, \label{webereq}\ee  This is Weber's equation, with solutions given by Weber's parabolic cylinder functions\cite{abram,nist,bateman,magnus}. These are real linear combinations of the  fundamental even and odd solutions which are regular as $x\rightarrow 0$,
\bea y_1(\kappa;x) & = & \Big[ 1- \kappa^2\, \frac{x^2}{2!}+ \Big(\kappa^4 - \frac{1}{2} \Big)\, \frac{x^4}{4!} + \cdots  \Big] \,\label{y1sol}\\
 y_2(\kappa;x) & = &x \,\Big[ 1- \kappa^2\, \frac{x^2}{3!}+ \Big(\kappa^4 - \frac{3}{2} \Big)\, \frac{x^4}{5!} + \cdots  \Big] \,,\label{y2sol}\eea and fulfill  the Wronskian condition
 \be \mathcal{W}_x \Big[y_2(\kappa;x),y_1(\kappa;x)\Big] =1 \,.\label{wronskianxy}\ee Weber's parabolic cylinder functions are given by\cite{abram}
 \be W[\kappa;\pm x] = \frac{1}{2^{3/4}}\,\Bigg[ \sqrt{\frac{G_1(\kappa)}{G_3(\kappa)}}\, y_1(\kappa,x) \mp \sqrt{\frac{2G_3(\kappa)}{G_1(\kappa)}}\, y_2(\kappa,x)\Bigg] \,,\label{parab}\ee with
 \be G_1(\kappa) = \Bigg|\Gamma\Big(\frac{1}{4} -i\frac{\kappa^2}{2} \Big)\Bigg|~~;~~ G_3(\kappa) = \Bigg|\Gamma\Big(\frac{3}{4}-i\frac{\kappa^2}{2}\Big) \Bigg|\,.\label{Gis2}\ee In particular for $\kappa=0$ ($k=0$, ``zero mode''),
  \be W[0,\pm x] = \frac{\sqrt{\pi x}}{2^{5/4}}\,\Bigg[J_{-1/4}\Big(\frac{x^2}{4} \Big) \pm J_{1/4}\Big(\frac{x^2}{4} \Big) \Bigg] \,,\label{zeromode} \ee with $J_{\nu}(z)$ the ordinary Bessel function.
  We seek  the  complex solutions $U_\kappa(x),U^*_\kappa(x)$ of the equation of motion (\ref{webereq}) obeying the Wronskian condition $\mathcal{W}_x[U_{\kappa}(x),U^*_{\kappa}(x)]=-i$. These are   related to the mode functions $\mathcal{U},\mathcal{U}^*$  that enter the expansion  (\ref{chi1exp}) as
  \be \mathcal{U}_k(\eta) \equiv \frac{U_{\kappa}(x)}{(2mH_R)^{1/4}}   \,.\label{Urelas}\ee
   It is straightforward to confirm that these are given by\cite{abram,nist,bateman,magnus}
 \bea U_\kappa(x) & = &  \frac{1}{ \sqrt{2}}\Bigg[\frac{W[\kappa,x]}{\sqrt{\mathcal{R}}} - i \sqrt{\mathcal{R}}\,\,W[\kappa,-x] \Bigg] \,,\label{U} \\
U^*_\kappa(x) & = &  \frac{1}{\sqrt{2}}\Bigg[\frac{W[\kappa,x]}{\sqrt{\mathcal{R}}} + i \sqrt{\mathcal{R}}\,W[\kappa,-x] \Bigg] \,,\label{Ucc}\eea where
\be \mathcal{R} = \sqrt{1+e^{-2\pi\kappa^2}}-e^{-\pi\kappa^2} \,.\label{R}\ee The reason for choosing the function $U_\kappa(x)$ is that it has a particularly important interpretation: for $\kappa^2 + x^2/4\gg 1$   when the Compton and/or (comoving) de Broglie wavelengths are much smaller than the Hubble radius their asymptotic behavior is\cite{abram,nist,bateman,magnus}
\be U_\kappa(x) =  \frac{e^{-i\int^x \sqrt{\kappa^2+\frac{{x'}^2}{4} }\,dx'}}{\sqrt{2\Big(\kappa^2+\frac{x^2}{4} \Big)^{1/2}}} \,,\label{asyU}\ee which coincides with the leading order Wentzel-Kramer-Brillouin (WKB) solution of the differential equation (\ref{webereq}).  This can be understood by writing the mode equation (\ref{webereq}) as
\be \Big[\frac{d^2}{dx^2} + \omega^2_{\kappa}(x)\Big]\,U_\kappa(x) =0 \,,\label{eqfreq} \ee for which a (WKB) ansatz\cite{birrell,parker} for its solution is
    \be U_\kappa(x) = \frac{e^{-i\int_0^x\,\Omega_{\kappa}(x')dx'}}{\sqrt{2\Omega_{\kappa}(x)}}\,,\label{adiadef} \ee
which when inserted into equation (\ref{eqfreq}) reveals that $\Omega_{\kappa}(x)$ must satisfy
     \be \Omega^2_{\kappa}(x) = \omega^2_\kappa(x)-\frac{1}{2}\Bigg[ \frac{\ddot{\Omega}_{\kappa}(x)}{\Omega_{\kappa}(x)} -\frac{3}{2}\frac{\dot{\Omega}^2_{\kappa}(x)}{\Omega^2_{\kappa}(x)}\Bigg]\,. \ee
The resulting equation can be solved in an \emph{adiabatic expansion}
    \be \Omega_{\kappa}(x) = \omega_{\kappa}(x) \Bigg[1-\frac{1}{2}\frac{\ddot{\omega}_{\kappa}(x)}{\omega^3_{\kappa}(x)}+\frac{3}{4}\Big(\frac{\dot{\omega}_{\kappa}(x)}{\omega^2_{\kappa}(x)}\Big)^2+\cdots \Bigg]\,, \label{adiaexp1}\ee in these expressions $\dot{\omega} \equiv d\omega/dx;\ddot{\omega}\equiv d^2 \omega /dx^2$, etc. The mode functions $U_\kappa(x)$ can be systematically obtained in this WKB asymptotic expansion, with the solution (\ref{asyU}) being the leading   order.

Using the relations (\ref{etaoft}, \ref{xvar} ),  and the scale factor (\ref{aofetarm}) for radiation domination, in terms of comoving time $t$ the solution (\ref{asyU}) becomes
\be U_k(t) = \frac{e^{-i\int^t E_k(t') \,dt'}}{\sqrt{2E_k(t)}} \,,\label{asyUoft}\ee where
\be E_k(t) = \sqrt{\frac{k^2}{a^2(t)}+m^2}\,,\label{energy} \ee is the energy measured by a local comoving observer. Therefore in this sub-Hubble limit, the wave functions describe adiabatic single particle states. In presence of cosmological expansion, these are the closest to Minkowski space-time mode functions describing single particle states, and embody the equivalence principle whereby deep inside the Hubble radius the wave functions behave similarly to Minkowski space time\cite{herring}.

In particular for the zero mode ( $\kappa=0$), it follows that
\be \omega_0(x) = \frac{x}{2}~~;~~\frac{\dot{\omega}_0(x)}{\omega^2_0}= \frac{2}{x^2} ~~;~~ \ddot{\omega}_0(x)=0\,,\label{largex}\ee therefore the leading order in the (WKB) expansion yields $\Omega_0(x) = x/2$ and  suppressing  the index $\kappa =0$ to simplify notation.  Therefore,  (from now on we focus solely on $\kappa=0$)  in the limit $x\gg 1$ and to leading order the mode function is given  by
\be U(x) =\frac{ e^{-i\frac{x^2}{4}}}{\sqrt{x}}\,,\label{Uwkb}\ee which is indeed the large $x$ behavior of the complex mode functions (\ref{U})\cite{abram,nist,bateman,magnus}.

The variable $x$ has an illuminating interpretation: from the
relations (\ref{hubble}), and the definition  (\ref{xvar}) it follows that
\be \frac{x^2}{2} =  \frac{m}{H(\eta)} =  \frac{d_H(\eta)}{\lambda_c} \label{hubrat} \ee where $d_H(\eta) = 1/H(\eta)$ is the Hubble radius, proportional to the particle horizon during radiation domination and $\lambda_c=1/m$ is the Compton wavelength of the particle. Therefore, for $x \ll 1$ the Compton wavelength is much larger than the Hubble radius, a situation to which we refer as super-Hubble. Furthermore, in terms of comoving time $t$ it follows from the above relations that
\be \frac{x^2}{4} = m t \,.\label{x2mt}\ee Therefore   the mode function $U(x)$ given by eqn. (\ref{Uwkb}) in the sub-Hubble limit $x\gg 1$ becomes
\be U(x) = \frac{e^{-imt}}{(4mt)^\frac{1}{4}}\,,\label{Uoft}\ee the exponential is precisely the Minkowski space time limit, whereas the denominator is a remnant of the cosmological expansion.  These aspects will be relevant to establish a relation with Minkowski space-time.

\vspace{1mm}

\section{Time evolution and expectation values}\label{sec:timeevol}
The action (\ref{conformalaction}) for the conformally rescaled fields in conformal time, yields the canonical momenta conjugate to the fields $\chi_{1,2}$ given by eqn. (\ref{canonical}), with the equal time commutation relations given by eqn. (\ref{commrel}), and the (conformal) Hamiltonian
\bea H(\eta) & = &  \int \Bigg\{ \frac{1}{2}\Big[ {\pi^2_1(\vx,\eta)} +  (\nabla \chi_1)^2 +\mathcal{M}^2_1(\eta)\chi^2_1(\vx,\eta)+   {\pi^2_2(\vx,\eta)} +   (\nabla \chi_2)^2 +\mathcal{M}^2_2(\eta)\chi^2_2(\vx,\eta)\Big] \nonumber \\ & + & \lambda \chi_1(\vx,\eta) :\chi^2_2(\vx,\eta): \Bigg\}\,d^3x\,. \label{conham}\eea  It is straightforward to show that the canonical commutation relations (\ref{commrel})  yield the Heisenberg field equations in conformal time as Hamiltonian equations of motion with the \emph{time dependent Hamiltonian} (\ref{conham}), and that these equations of motion are the same as those obtained from the variation of the action (\ref{conformalaction}). Therefore, the conformal Hamiltonian (\ref{conham}) is the generator of time evolution. Hence for any operator $\mathcal{O}$ which is time independent in the Schroedinger picture, Heisenberg's Hamilton's equation of motion become
\be \frac{d}{d\eta}\mathcal{O}(\vx,\eta) = i \big[ H(\eta), \mathcal{O}(\vx,\eta)\big] \,,\label{Heiseom}\ee whose solution is
\be \mathcal{O}(\vx,\eta_0) = U^{-1}(\eta,\eta_0)\,\mathcal{O}(\vx,\eta_0)\,U(\eta,\eta_0) \,,\label{solhem}\ee where the unitary time evolution operator is given by
\be U(\eta,\eta_0)= T\Big( e^{-i\int^\eta_{\eta_0}\,H(\eta^{'})\,d\eta^{'}} \Big)~~;~~ U^{-1}(\eta,\eta_0) = \widetilde{T}\Big( e^{i\int^\eta_{\eta_0}\,H(\eta^{'})\,d\eta^{'}} \Big) \,,\label{Uop}\ee  and $T,\widetilde{T}$ are the time ordering and anti-ordering symbols respectively. With an initial state described by a density matrix $\rho(\eta_0)$, normalized such that $\mathrm{Tr} \rho(\eta_0) =1$, expectation values of a Heisenberg field operator are given by
\be \langle \mathcal{O}(\eta) \rangle = \mathrm{Tr}\mathcal{O}(\eta) \,\rho(\eta_0) \,.\label{exval}\ee  Expectation values and correlation functions   are obtained via functional derivatives of the generating functional\cite{beilok,boylee}
\be \mathcal{Z}[J^+,J^-] \equiv \mathrm{Tr}\Big[ U(\eta,\eta_0;J^+)\,\rho(\eta_0)\,U^{-1}(\eta,\eta_0;J^-)\Big] \,,\label{ZJ}\ee with respect to the external sources $J^\pm$, where
\be   U(\eta,\eta_0;J^+) = \mathrm{T} \Big( e^{-i \int^\eta_{\eta_0}H(\eta';J^+)} \Big)~~;~~  U^{-1}(\eta,\eta_0;J^-)= \widetilde{T}\Big( e^{i\int^\eta_{\eta_0}\,H(\eta^{'};J^-)\,d\eta^{'}} \Big)\ee  with the Hamiltonian including the coupling to the external sources
\be H(\eta,J^{\pm}) \equiv H(\eta) + \int d^3 x J^{\pm}(\vx,\eta)\,\mathcal{O}(\vx,\eta) \,.\label{Hjs}\ee For example
\bea \langle   \mathcal{O}^+(\vx_1,\eta_1)\mathcal{O}^+(\vx_2,\eta_2) \rangle & \equiv & \langle T \Big(\mathcal{O}(\vx_1,\eta_1)\mathcal{O}(\vx_2,\eta_2)\Big)\rangle  =   \mathrm{Tr}  \Big(T \mathcal{O}(\vx_1,\eta_1)\mathcal{O}(\vx_2,\eta_2)\Big) \rho(\eta_0) \nonumber \\ & = &  -\frac{\delta^2\,\mathcal{Z}[J^+,J^-]}{ \delta J^+(\vx_1,\eta_1)\delta J^+(\vx_2,\eta_2) } \Big|_{J^+=J^-=0}\,,  \label{timor} \eea
\bea
\langle \mathcal{O}^-(\vx_2,\eta_2)\mathcal{O}^+(\vx_1,\eta_1)\rangle & \equiv & \langle \mathcal{O}(\vx_2,\eta_2)\mathcal{O}(\vx_1,\eta_1)\rangle =  \mathrm{Tr}      \mathcal{O}(\vx_1,\eta_1)\, \rho(\eta_0)\,\mathcal{O}(\vx_2,\eta_2) \nonumber \\ & = &  \frac{\delta^2\,\mathcal{Z}[J^+,J^-]}{ \delta J^+(\vx_1,\eta_1)\delta J^-(\vx_2,\eta_2) }\Big|_{J^+=J^-=0}\,, \label{timo}\eea etc. An important result is that
\bea  && \langle  \mathcal{O}^+(\vx,\eta)\rangle   \equiv     \mathrm{Tr}      \mathcal{O}(\vx,\eta)\, \rho(\eta_0)  = -i\frac{\delta\,\mathcal{Z}[J^+,J^-]}{ \delta J^+(\vx,\eta)} \Big|_{J^+=J^-=0} \nonumber \\ & = &  \langle  \mathcal{O}^-(\vx,\eta)\rangle   \equiv    \mathrm{Tr}       \rho(\eta_0) \,  \mathcal{O}(\vx,\eta)= i\frac{\delta\,\mathcal{Z}[J^+,J^-]}{ \delta J^-(\vx,\eta)} \Big|_{J^+=J^-=0} \,. \label{exvalj}\eea

 Referring  to the fields $\chi_{1,2}$ collectively as $\chi$, the generating functional (\ref{ZJ}) in the field representation can be written in a functional integral representation
\be \mathcal{Z}[J^+,J^-] = \int D\chi_f D\chi_i D\chi'_i \,\bra{\chi_f} U(\eta,\eta_0;J^+) \ket{\chi_i}\,\bra{\chi_i}\rho(\eta_0)\ket{\chi'_i}\bra{\chi'_i}U^{-1}(\eta,\eta_0;J^-)\ket{\chi_f}\,,\label{parti} \ee in turn the field matrix elements of the evolution operators can be written as path integrals, namely
\bea  \bra{\chi_f} U(\eta,\eta_0;J^+) \ket{\chi_i} & \equiv &  \int \mathcal{D}\chi^+ \,e^{i\int  \mathcal{L}[\chi^+;J^+]\,  d^4x} ~~;~~ \chi^+(\eta_0) = \chi_i;\chi^+(\eta)=\chi_f \,,\label{lplus} \\ \bra{\chi'_i} U^{-1}(\eta,\eta_0;J^-) \ket{\chi_f} & \equiv &  \int \mathcal{D}\chi^- \,e^{-i\int \mathcal{L}[\chi^-;J^-]\,  d^4x} ~~;~~ \chi^-(\eta_0) = \chi'_i;\chi^-(\eta)=\chi_f \,,\label{lmin}\eea where
\be \mathcal{L}[\chi^\pm;J^\pm] =\biggl\{\sum_{j=1,2}\Bigl[\tfrac{1}{2}  \,\Big(\frac{\partial \chi^\pm_j}{\partial \eta} \Big)^2 -\tfrac{1}{2} \,\bigl( \nabla \chi^\pm_j \bigr)^2 -\tfrac{1}{2} {\chi^\pm}^2_j\, \mathcal{M}^2_j(\eta)-J^\pm_j \,\chi^\pm_j \Bigr] -   \lambda \,a(\eta)\,\chi^\pm_1 \,:(\chi^\pm_2)^2: \ \,  \biggr\}\,.\label{lagspm}  \ee Finally, the functional and path integral representation of the generating functional becomes
\be \mathcal{Z}[J^+,J^-] =   \int D\chi_f D\chi_i D\chi'_i \int \mathcal{D}\chi^+ \mathcal{D}\chi^- e^{i \int \Big[\mathcal{L}[\chi^+;J^+]-\mathcal{L}[\chi^-;J^-]\Big] d^4x} \,\rho(\chi_i,\chi'_i;\eta_0) \,,\label{pathint} \ee with the boundary conditions on the fields $\chi^\pm$ given by eqns. (\ref{lplus},\ref{lmin}) and the notation $\int d^4x \equiv \int^\eta_{\eta_0} d\eta' \int d^3 x$. This is the in-in or Schwinger-Keldysh formulation of non-equilibrium quantum field theory\cite{schwinger,keldysh,maha,jordan}.

Our objective is to obtain the linerized equation of motion for the   expectation value of the massive scalar field $\chi_1$ (or $\phi_1=\chi_1/a(\eta)$), namely
\be \mathrm{Tr} \, \chi_1(\vx,\eta) \rho(\eta_0) \equiv \X(\eta) \,,\label{Xv}\ee  where we consider $\X$ to be spatially homogeneous, consistently with homogeneity and isotropy of the cosmology, hence only the zero momentum component of $\chi_1$ acquires an expectation value. The equation of motion for $\X$ is obtained by following the identity (\ref{exvalj}) which implies that $\langle \chi^+ \rangle = \langle \chi^- \rangle = \X$.

Writing
\be \chi^\pm_1(\vx,\eta) = \X(\eta) + \delta^\pm(\vx,\eta)\,, \label{shift}\ee    in the Lagrangian $\mathcal{L}[\chi^\pm,J^\pm]$ in equation (\ref{lagspm}) and requesting that
\be \langle \delta^\pm(\vx,\eta)\rangle =0\,,\label{zerodelta}\ee  to all orders in perturbation theory. Upon integration by parts and neglecting surface terms, and coupling sources only to the fluctuating fields $\chi^\pm_2,\delta^\pm$,  we obtain (primes denote $\partial/\partial \eta$)
\bea && i\int  \Big[\mathcal{L}[\X;\delta^+,\chi^+_2,J^+] - \mathcal{L}[\X;\delta^-,\chi^-_2,J^-]\Big]  d^4x = \nonumber \\ && -i\int  \Big[\X^{''}(\eta)+\M^2_1(\eta)\X(\eta)\Big]\delta^+(\vx,\eta)\,d^4 x -  \Big( \delta^+ \rightarrow \delta^- \Big)\nonumber \\
&& + i \int \frac{1}{2}\,\Bigg\{\Big(\frac{\partial \delta^+}{\partial \eta}\Big)^2-\Big( \nabla \delta^+ \Big)^2-\mathcal{M}^2_1(\eta)\,{\delta^+}^2+ J^+_1\delta^+ + \Big(\frac{\partial \chi^+_2}{\partial \eta}\Big)^2-\Big( \nabla \chi^+_2 \Big)^2 + J^+_2\chi^+_2\Bigg\}\,d^4 x \nonumber \\
&& -i \int \frac{1}{2}\,\Bigg\{\Big(\frac{\partial \delta^-}{\partial \eta}\Big)^2-\Big( \nabla \delta^- \Big)^2-\mathcal{M}^2_1(\eta)\,{\delta^-}^2+ J^-_1\delta^- +\Big(\frac{\partial \chi^-_2}{\partial \eta}\Big)^2-\Big( \nabla \chi^-_2 \Big)^2 + J^-_2\chi^-_2\Bigg\}\,d^4 x \nonumber \\
&&-i\lambda \int a(\eta)\,\Bigg\{\Big(\X+\delta^+\Big)\,  :(\chi^+_2)^2:- \Big(\X+\delta^-\Big)\,  :(\chi^-_2)^2:\Bigg\}\,d^4x \,.\label{shiftedlag}
 \eea

       \begin{figure}[ht!]
\begin{center}
\includegraphics[height=3in,width=4in,keepaspectratio=true]{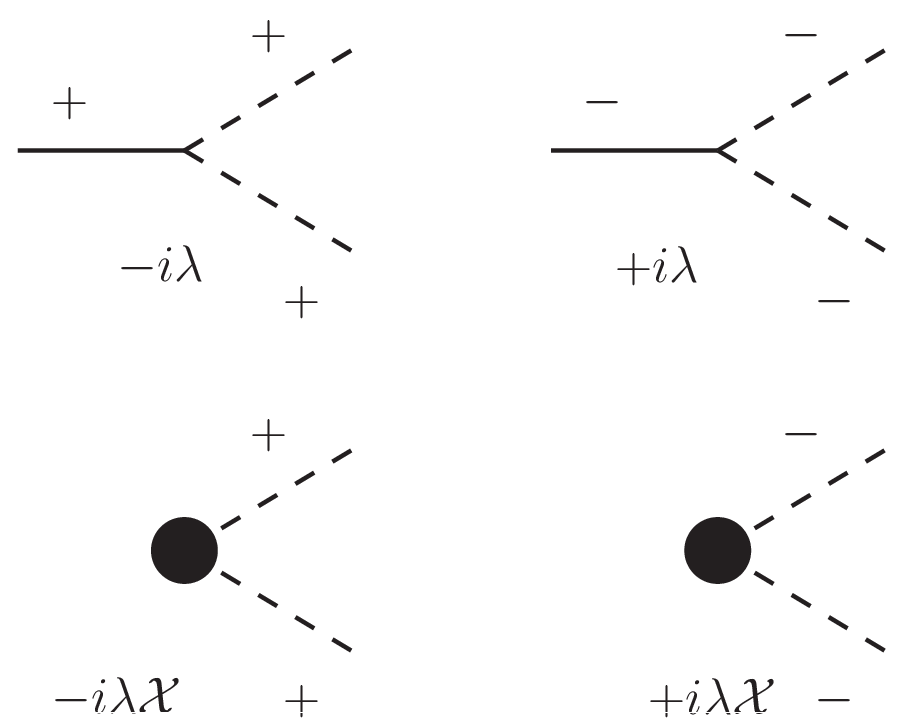}
\caption{Interaction vertices from the Lagrangian (\ref{shiftedlag}). The solid lines correspond to the fluctuations of the massive field $\delta^\pm$ the dashed lines to the massless fields $\chi^\pm_2$. The dark dot stands for the condensate $\X$ . }
\label{fig:vertices}
\end{center}
\end{figure}

 Following the method introduced in ref.\cite{boytad}, the equation of motion for $\X$ is obtained by treating the \emph{first} and last terms on the right hand side of (\ref{shiftedlag}) as perturbations in the free field theory defined by the quadratic terms in fluctuations $\delta^\pm; \chi^\pm_2$, namely the second and third lines on the right hand side of equation (\ref{shiftedlag}).  For example, to zeroth order in the coupling $\lambda$,
 \be \langle \delta^+(\vec{y},\eta')\rangle  = -i \int \Bigg\{ \Big[\langle \delta^+(\vec{y},\eta')\delta^+(\vec{x},\eta)\rangle -\langle \delta^+(\vec{y},\eta')\delta^-(\vec{x},\eta)\rangle\Big]\Big[\X^{''}(\eta)+\M^2_1(\eta)\X(\eta) \Big]\Bigg\}\,d^4x=0 \label{zerothord} \,,  \ee since the correlation functions
 $\langle \delta^+(\vec{y},\eta')\delta^\pm(\vec{x},\eta)\rangle$ are independent,
  it follows that
 \be \X^{''}(\eta)+\M^2_1(\eta)\X(\eta) =0 \,,\label{zerotheom}\ee the same equation of motion is obtained from $\langle \delta^-(\vec{y},\eta')\rangle =0$. Because perturbation theory is carried out in the free field theory of the $\delta^\pm,\chi^\pm_2$ fields, these must always appear in pairs in correlation functions, therefore the next contribution is of $\mathcal{O}(\lambda^2)$ corresponding to a one-loop self-energy, which is given by
 \bea && \langle \delta^+(\vec{y},\eta_1)\rangle^{(1\,\mathrm{loop})}   =    -\lambda^2\,\int d^4x \langle \delta^+(\vec{y},\eta_1)\delta^+(\vec{x},\eta)\rangle\,a(\eta)\times \label{1lup} \\ & & \int d^4x' \,a(\eta')\,\Big[\langle :(\chi^+_2(\vx,\eta))^2:\,:(\chi^+_2(\vx',\eta'))^2:\rangle-  \langle :(\chi^+_2(\vx,\eta))^2:\,:(\chi^-_2(\vx',\eta'))^2:\rangle\Big]\X(\eta') \,d^4x'\,,\nonumber\eea where we have only considered the contribution from the correlation function $\langle \delta^+(\vec{y},\eta_1)\delta^+(\vec{x},\eta)\rangle\ $ since the other yields the same equation, consistently with the zeroth order case.

     \begin{figure}[ht!]
\begin{center}
\includegraphics[height=3in,width=4in,keepaspectratio=true]{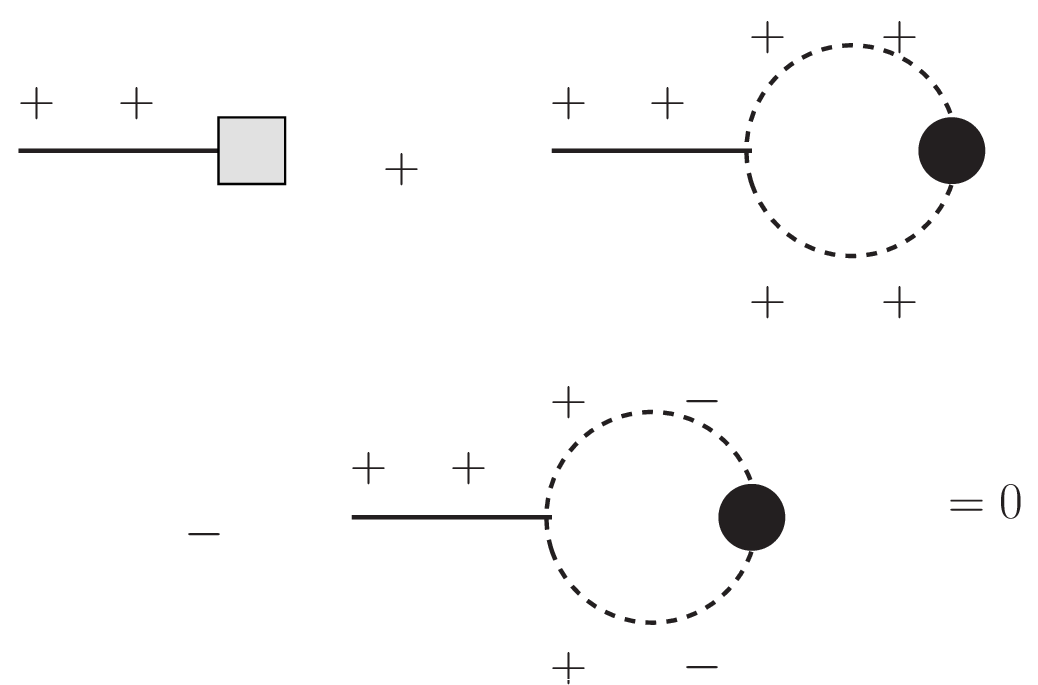}
\caption{Representation of the diagrams leading to the equation of motion (\ref{1lupeom}). The solid line is the propagator $\langle \delta^+ \delta^+\rangle$, the square is the tree level term $\X''+\mathcal{M}^2_1\,\X$, the dark dot stands for $\X$, the dashed lines correspond to the massless propagators with the corresponding branches $\pm$. The loops represent the self energy $\Sigma$.}
\label{fig:eom}
\end{center}
\end{figure}

  Combining the zeroth order term (\ref{zerothord}) and the one loop contribution yields the linearized equation of motion up to one loop order
  \be \X^{''}(\eta)+\M^2_1(\eta)\X(\eta)+\int \Sigma(\vx,\vx';\eta,\eta')\,\X(\eta')\,d^4x' =0 \,,\label{1lupeom}\ee where the self-energy, depicted by the loops in fig. (\ref{fig:eom}),  is given by
  \be \Sigma(\vx,\vx';\eta,\eta') = -i\lambda^2\,a(\eta)  \,a(\eta')\,\Big[\langle :(\chi^+_2(\vx,\eta))^2:\,:(\chi^+_2(\vx',\eta'))^2:\rangle-  \langle :(\chi^+_2(\vx,\eta))^2:\,:(\chi^-_2(\vx',\eta'))^2:\rangle\Big]\,.\label{sigma}  \ee  Using Wick's theorem, we find
  \bea && \langle :(\chi^+_2(\vx,\eta))^2:\,:(\chi^+_2(\vx',\eta'))^2:\rangle   =   2\,\Big[ G^>(\vx,\vx';\eta,\eta')\Theta(\eta-\eta')+ G^<(\vx,\vx';\eta,\eta')\Theta(\eta'-\eta)\Big]\, \nonumber \\ && \langle :(\chi^+_2(\vx,\eta))^2:\,:(\chi^-_2(\vx',\eta'))^2:\rangle   = 2\, G^<(\vx,\vx';\eta,\eta')\,,\label{Gis}\eea where
  \bea  G^>(\vx,\vx';\eta,\eta') & = &  \langle    \chi_2(\vx,\eta) \, \chi_2(\vx',\eta') \rangle\,\langle \chi_2(\vx,\eta) \, \chi_2(\vx',\eta') \rangle \,\label{gplu}\\
  G^<(\vx,\vx';\eta,\eta') & = &  \langle \chi_2(\vx',\eta') \,\chi_2(\vx,\eta)\rangle\,\langle  \chi_2(\vx',\eta') \,\chi_2(\vx,\eta) \rangle \,, \label{gmin}\eea since $\chi_2$ is a real scalar field, it follows that
  \be G^<(\vx,\vx';\eta,\eta') = G^>(\vx',\vx;\eta',\eta)\,. \label{iden}\ee

  Therefore,
  \be  \Sigma(\vx,\vx';\eta,\eta') = -2i\lambda^2\,a(\eta)  \,a(\eta')\,\Big[G^>(\vx,\vx';\eta,\eta')-[G^<(\vx,\vx';\eta,\eta')  \Big]\Theta(\eta-\eta')\,,\label{retsig}\ee this is the \emph{retarded} self energy. The in-in (or Schwinger-Keldysh) formulation of non-equilibrium quantum field theory yields causal (retarded) equations of motion in terms of a non-local retarded self-energy kernel.

 We will consider an initial density matrix describing an equilibrium  thermal bath at temperature $T$ for the massless particles $\chi_2$,
 \be \rho(\eta_0) \propto e^{-H_0(\chi_2)/T} \,, \label{rhoin2}\ee where $H_0(\chi_2)$ is the free field part of the conformal time Hamiltonian (\ref{conham}) for the massless field $\chi_2$. Therefore with the expansion (\ref{chi2exp}) of the quantized $\chi_2$ field, we find
 \be \langle    \chi_2(\vx,\eta) \, \chi_2(\vx',\eta') \rangle = \frac{1}{V}\sum_{\vk}\frac{1}{2k} \Big[(1+n(k))\,e^{-ik(\eta-\eta')} + n(k)\, e^{ik(\eta-\eta')} \Big]\,\,e^{i\vk\cdot(\vx-\vx')}\,,\label{thermalcorr}\ee with the thermal distribution function
 \be n(k) = \frac{1}{e^{\frac{k}{T}}-1}\,.\label{nofk}\ee

  A this point, it proves illuminating to write the two point correlation function of the original, unscaled field $\phi_2$ in \emph{comoving time}, namely
  \be \langle \phi_2(\vx,t)\phi_2(\vx',t')\rangle = \frac{1}{2H_R\,V\,\sqrt{t\,t'}}\sum_{\vk}\frac{1}{2k} \Big[(1+n(k))\,e^{-ik\sqrt{2H_R}(\sqrt{t}-\sqrt{t'})} + n(k)\, e^{ik\sqrt{2H_R}(\sqrt{t}-\sqrt{t'})} \Big]\,\,e^{i\vk\cdot(\vx-\vx')} \,.\label{2ptcomo}\ee While this correlation function features spatial translational invariance, it does \emph{not} feature time translational invariance, this is a direct manifestation of the lack of a time-like Killing vector of the (FRW) cosmology. We note that whereas the two point funtion of the \emph{massless} field $\chi_2$  is translational invariant in conformal time, the self-energy (\ref{retsig}) is not. In particular, this lack of time translational invariance prevents a spectral (Callen-Lehmann) representation of the self-energy, which is available in Minkowski space time (see appendix (\ref{app:eomlap})).

  Because the expectation value $\X(\eta)$ is homogeneous and does not depend on the spatial coordinate, the spatial integral in $d^3x'$ in equation (\ref{1lupeom}) can be done explicitly, yielding the equation of motion for the homogeneous condensate
 \be \X^{''}(\eta)+m^2 H^2_R \, \eta^2 \X(\eta)+\int^\eta_{\eta_0} \Sigma(\eta,\eta')\,\X(\eta')\,d\eta' =0\,,\label{fineom} \ee where we have kept the same symbol for the self-energy to simplify notation. A straightforward calculation after the spatial integration yields the one-loop self energy
 \be \Sigma(\eta,\eta') = -2i \lambda^2\,a(\eta)\,a(\eta') I(\eta-\eta')\,,\label{sigi}\ee where
 \be I(\eta-\eta') \equiv I_0(\eta-\eta')+ I_T(\eta-\eta') = \int \frac{1}{4k^2}\,[1+2\,n(k)]\,\Big[e^{-2ik(\eta-\eta')}-e^{2ik(\eta-\eta')}  \Big] \,\frac{d^3k}{(2\pi)^3} \,,\label{IntI}\ee where $I_0,I_T$ correspond to the zero and finite temperature components respectively.
 The factor $(1+2 n(k))$ has an important intepretation: it arises from the difference between the process of stimulated emission and recombination (or absorption) of massless quanta in the medium: the stimulated emission (decay) $\chi_1(\vk=0) \rightarrow \chi_2(\vk) \chi_2(-\vk)$ contributes with the factor $(1+n(k))(1+n(k))$ whereas the recombination from the massless quanta populated in the medium $\chi_2(\vk) \chi_2(-\vk) \rightarrow \chi_1(\vk=0)$ contributes with the factor $n(k)n(k)$ (from annihilation of a particle with momentum $\vk$ and another with $-\vk$ in the medium). Therefore, the finite temperature contribution to the self-energy is a \emph{real-time} manifestation of the balance between stimulated emission and absorption (recombination).

 We note, again,  that whereas $I(\eta-\eta')$ is translational invariant in \emph{conformal time}, the full self energy (\ref{sigi}) is not, again preventing a spectral representation in the frequency domain. This is important because such a spectral representation of the self-energy has been used in the literature\cite{berera}, as the basis for describing dissipative processes via a local friction term.

 The integral for the  zero temperature contribution (the ``1'' in eqn. (\ref{IntI})), is carried out by introducing a convergence factor $\pm i\epsilon, \epsilon \rightarrow 0^+$ in the exponents, with the result
 \be I_0(\eta-\eta') = -\frac{i}{8\pi^2} \frac{(\eta-\eta')}{(\eta-\eta')^2+\epsilon^2}~~;~~ \epsilon \rightarrow 0^+\,,\label{I0}\ee which reflects an ultraviolet short distance singularity as $\eta' \rightarrow \eta$. This singularity is precisely the ultraviolet singularity of the one loop diagram and agrees with the short distance behavior obtained from the operator product expansion in   conformal time.

 The finite temperature contribution $I_T(\eta-\eta')$ is obtained by expanding
 \be \frac{1}{e^{\frac{k}{T}}-1}= \sum_{l=1}^{\infty} e^{-lk/T} \,,\label{exp}\ee and integrating term by term, yielding
 \be I_T(\eta-\eta') =  -\frac{2i}{8\pi^2}\,\sum_{l=1}^{\infty} \frac{(\eta-\eta')}{\Big( \frac{l}{2T}\Big)^2+\Big( \eta-\eta'\Big)^2}\,, \label{IT}\ee the identity (\ref{ident}) in appendix (\ref{app:subTnonzero}) allows us to provide the  more compact form for this contribution
 \be I_T(\eta-\eta') = -\frac{i}{8\pi^2}\,(2\pi T)\,\Big[ \coth\big[2\pi T (\eta-\eta')\big]   - \frac{1}{2\pi T (\eta-\eta')}\Big]\,.\label{ITfin}  \ee We note that  the limit
 \be {\coth[z]-\frac{1}{z}}_{~~ \overrightarrow{z\longrightarrow 0}~~ } \propto z \,,\label{shortd}\ee indicates, as expected, that the finite temperature contribution is ultraviolet finite. Therefore the ultraviolet divergence of the one loop self energy is solely determined by the zero temperature contribution. Combining the zero and finite temperature results, we finally obtain
 \be \Sigma(\eta,\eta') = -g^2\,a(\eta)\,a(\eta')\Bigg\{\frac{(\eta-\eta')}{(\eta-\eta')^2+\epsilon^2} +  (2\pi T)\,\Big[ \coth\big[2\pi T (\eta-\eta')\big]   - \frac{1}{2\pi T (\eta-\eta')}\Big]  \Bigg\}    ~~;~~\epsilon\rightarrow 0^+\,.\label{sigtot}\ee where
  \be g = \frac{\lambda}{2\pi} \,. \label{coupli}\ee

 The self-energy contribution to the equation of motion, cannot be written as a local ``friction'' type term. In fact as explicitly displayed by eqn. (\ref{sigtot}) the self-energy kernel is non-local, retarded,  and features long range memory as is explicit in the zero temperature contribution that falls-off as $1/(\eta -\eta')$ as $\eta-\eta' \rightarrow \infty$, a hallmark of massless particles.

 To make this distinction with the phenomenological friction term more explicit, consider the typical phenomenological equation often used in the cosmological literature (see ref.\cite{kolb,berera} and references therein)
 \be \ddot{\phi}(t)  + 3 H(t) \dot{\phi}(t) + m^2 \phi(t) + \Gamma \dot{\phi}(t) =0 \,,\label{fric}\ee with $\Gamma$ the Minkowsky decay rate, which for the case under consideration is $\Gamma = \pi g^2/2m$. Passing to conformal time, with the conformal rescaling of the field $\phi(t) = \X(\eta)/a(\eta)$,   we obtain  with the notation $'\equiv \frac{d}{d\eta}$,
 \be \X^{''}(\eta)+m^2 H^2_R \, \eta^2 \X(\eta)+\Gamma\,H_R\,\eta^2\,\Big(\frac{\X(\eta)}{\eta}\Big)'=0 \,,\label{friceq} \ee from which it is clear that a local friction term $\Gamma\,\dot{\phi}$ \emph{cannot} describe the non-local self-energy contribution.

 Precisely because of the scale factors $a(\eta)$ in the equation of motion and the self-energy \emph{there is no} time translational invariance as in Minkowski space time, reflecting the lack of a time-like Killing vector of the FRW metric. The Minkowski space-time limit is obtained by setting $a(\eta) \rightarrow 1$ and $\eta,\eta' \rightarrow t,t'$, a relation that allows us to directly compare with the Minkowski space time result. Such comparison is presented in detail in   appendices  (\ref{app:eomlap},\ref{app:drg}).

 Because of this  lack of time translational invariance, the equation of motion (\ref{fineom})  while linear,  unlike the case of Minkowski space-time, it cannot be solved either by Fourier or Laplace transforms, requiring a different framework alternative to the usual approaches in flat space-time.

 We now compare the solutions of the non-local equation (\ref{fineom}) with the self-energy given by (\ref{sigtot})  with that of the local equation with the phenomenological friction term (\ref{friceq}).

 \vspace{1mm}

 \section{Dynamical Renormalization Group}\label{sec:drg}

 Anticipating mass renormalization as a consequence of the ultraviolet divergence of the self energy we write
 \be m^2 = m^2_R + \delta m^2 \,,\label{renmass}\ee where $\delta m^2$ is a counterterm that will be fixed by requesting it to cancel the ultraviolet divergences order by order in perturbation theory. We change variables as in eqn. (\ref{xvar}) but in terms of the renormalized mass
 namely\footnote{We kept the same name for the dimensionless variable to simplify notation.},
 \be x = \sqrt{2 m_R H_R}~\eta \label{nux}\ee  taking the lower limit $\eta_0 \rightarrow 0$ and introducing the dimensionless variables
 \be  \Delta_m   =  \frac{\delta m^2}{4m^2_R} ~~;~~ \widetilde{g}^2 = \frac{g^2}{4m^2_R} = \frac{\lambda^2}{8\pi\,m^2_R}~~;~~ \widetilde{T}  =   \frac{T}{\sqrt{2 m_R H_R}} ~~;~~ \widetilde{\epsilon}  =   \frac{\epsilon}{\sqrt{2 m_R H_R}}\,,\label{nudimvars}
\ee
   the equation of motion (\ref{fineom}) becomes
\be \frac{d^2}{dx^2} \X(x)+ \frac{x^2}{4}\,\X(x) + \Delta_m \, x^2 \X(x) + \widetilde{g}^2\,\int^x_0 \widetilde{\Sigma}(x,x')\,\X(x')\,dx' =0 \,,\label{eomofx} \ee where
  \be \widetilde{\Sigma}(x,x') = - \,x\,x'\Bigg\{\frac{(x-x')}{(x-x')^2+\widetilde{\epsilon}^2} +  (2\pi \widetilde{T})\,\Big[ \coth\big[2\pi \widetilde{T} (x-x')\big]   - \frac{1}{2\pi \widetilde{T} (x-x')}\Big]  \Bigg\}\,.\label{tilsig}\ee
  The strategy is to solve the equation of motion (\ref{eomofx}) in perturbation theory in a power series in the dimensionless coupling $\widetilde{g}$,
  \be \X(x) = \X^{(0)}(x)+\widetilde{g}^2\,\X^{(1)}(x)+\cdots \label{Xseries}\ee
 The perturbative solution of the equations of motion yield secular terms, namely terms that grow in time. The dynamical renormalization group, originally developed to resum these secular terms within the theory of amplitude equations for pattern formation\cite{drggold}, provides a non-perturbative and systematic framework that   yields amplitude equations with improved and controlled asymptotics. This method  has been extended to study   processes in non-equilibrium quantum field theory for example inflaton decay\cite{drg1} and relaxation including anomalous decay in \cite{drg}      by providing a systematic resummation of the perturbative solution of the equations of motion for expectation values that is asymptotically convergent. The explicit example of massive scalar decay into massless fields in Minkowski space-time is provided in detail in appendix (\ref{app:drg}) yielding a direct comparison with the case of radiation dominated cosmology. Appendix (\ref{app:eomlap}) presents the usual approach to obtain the time evolution of the amplitude of the condensate from the Fourier-Laplace transform of the Dyson-resummed propagator including self-energy radiative corrections in Minkowski space-time, and appendix (\ref{app:drg}) provides a detailed analysis of this process based on the dynamical renormalization group, again in Minkowski space-time for comparison.

 As a first step in this program, the counterterm $\Delta_m$ will be adjusted order by order in perturbation theory to cancel the ultraviolet divergences from the self-energy corrections. By writing
  \be \Delta_m = \widetilde{g}^2\,\Delta^{(1)}_m + \cdots \,,\label{count}\ee we obtain the hierarchy of equations for the terms in the series
  \bea \frac{d^2}{dx^2} \X^{(0)}(x)+ \frac{x^2}{4}\,\X^{(0)}(x) & = & 0 \,,\label{zerote}\\
\frac{d^2}{dx^2} \X^{(1)}(x)+ \frac{x^2}{4}\,\X^{(1)}(x) & = & S(x)\,,\label{firste}\\
\vdots ~~~~~~ & = &  ~~~~~~\vdots \nonumber \eea where the source term
\be S(x) = -\Big[ \Delta^{(1)}_m \, x^2 \X^{(0)}(x) +  \int^x_0 \widetilde{\Sigma}(x,x')\,\X^{(0)}(x')\,dx'   \Big]\,.\label{source}\ee

In the following analysis, after using the mass counterterm $\Delta^{(1)}_m$ to cancel the ultraviolet divergences,  we refer to the renormalized mass with the same symbol,  $m$,  to simplify notation.

The complicated analytic form of the mode functions $U_{\kappa}(x)$ prevents us from providing an analytic treatment all throughout the time evolution. Therefore, to make progress we analyze separately the super-Hubble limit $x\ll 1$ and the sub-Hubble limit $x\gg1$. This will allow us to illuminate the important aspects of space time curvature in the decay dynamics of the condensate and to compare directly with the Minkowski space-time limit studied in detail in appendices (\ref{app:eomlap},\ref{app:drg}). We   discuss the ``matching'' between the two regimes  in   section (\ref{sec:discussion}).

\subsection{Super-Hubble limit}\label{sub:super}
As discussed above, the case when the Compton wavelength of the massive particle $1/m$ is much larger than the Hubble radius corresponds to $m/H(\eta)\ll 1$, namely $x \ll 1$ (see eqn. (\ref{hubrat})), we refer to this as the super-Hubble limit. In this limit it is convenient to write the solution of the zeroth order equation, $\X^{(0)}(x)$ as a linear combination of the real even and odd functions (see equations (\ref{y1sol},\ref{y2sol}) for $\kappa =0$, namely $k=0$)
\bea y_e(x) & = & \Big[ 1 - \frac{1}{2}  \, \frac{x^4}{4!} + \cdots  \Big] \,\label{yeven}\\
 y_o(x) & = &x \,\Big[ 1  - \frac{3}{2}  \, \frac{x^4}{5!} + \cdots  \Big] \,,\label{yodd}\eea with Wronskian
 \be \mathcal{W}_x[y_o(x),y_e(x)] = 1\,, \label{Weo}\ee hence
 \be \X^{(0)}(x) = \widetilde{A}\,y_o(x) +
 \widetilde{B}\,y_e(x) \,,\label{oecombo}\ee where the coefficients are determined by initial conditions. The conformal rescaling of the original field, equation (\ref{rescale}) implies that the homogeneous component of the original massive field $\phi_1$  is related\footnote{Do not confuse the coordinate vector $\vx$ with the dimensionless variable $x$. When referring to coordinates $\vx$ is always a vector, the homogeneous component does not depend on the spatial coordinates. } to $\X(x)$ as $\phi_1(\eta) = \X(x(\eta))/a(\eta)$. With the scale factor (\ref{aofetarm}) and the definition  (\ref{xvar}),   the homogeneous condensate solution of the original $\phi_1$ field to zeroth order is given by
 \be \phi_1(t(\eta)) \propto \frac{\X(x)}{x} \,.\label{relafi}\ee

  As an initial condition,  we request that  the amplitude of the $\phi_1$ condensate  be regular as $\eta \rightarrow 0$, namely the beginning of the radiation dominated stage, (corresponding to $x \rightarrow 0$). This requires the coefficient $\widetilde{B}=0$.   With the relation (\ref{zeromode}),  it follows that $\X^{(0)}(x)$ is the linear combination   that yields
  \be \phi^{(0)}_1(t) =  \mathcal{A}\,\,\frac{(mt)^{\frac{1}{4}}}{a(t)}  \, J_{\frac{1}{4}}(mt)\,,\label{fibes}\ee
   with $\mathcal{A}$ a constant real amplitude. As $mt\rightarrow 0$ it follows that $\phi^{(0)}_1(t) \rightarrow \mathrm{constant}$, namely the homogeneous condensate does not feature oscillations in the super-Hubble limit, and begins to oscillate when the Compton wavelength becomes of the same order as the Hubble radius, namely $mt \simeq 1$, in other words, when the Compton wavelength enters the particle horizon. This is the usual assumed behavior of the condensate.

   Therefore we consider the zeroth order solution
   \be \X^{(0)}(x) = \widetilde{A} \,x\,V(x)~~;~~V(x) = \Big[ 1 - \frac{x^4}{80}+\cdots\Big]\,, \label{chizi}\ee where we have written
   the odd solution (\ref{yodd}) as $y_o(x) \equiv x\,V(x)$ to highlight that the corrections to the linear term are of $\mathcal{O}(x^5)$ and highly suppressed in the super-Hubble limit $x\ll 1$.

\textbf{Zero Temperature:}
We will first analyze the zero temperature contribution to the self energy, namely
\be \widetilde{\Sigma}_0(x,x') = - \,x\,x'\,\,\frac{(x-x')}{(x-x')^2+\widetilde{\epsilon}^2}=  x\,x' \,\frac{1}{2}\,\frac{d}{dx'}\, \ln\Big[(x-x')^2+ \widetilde{\epsilon}^2\Big]\,,\label{sigze} \ee which when inserted in the integral in eqn. (\ref{firste}) yields
\be \int^x_0 \widetilde{\Sigma}(x,x')\,\X^{(0)}(x')\,dx'  = x^2\,\X^{(0)}(x)\,\ln(\widetilde{\epsilon})- x\,\int^x_0 \ln[x-x'] \,\frac{d}{dx'}\Big[ x'\,\X^{(0)}(x')\Big]\,dx' \,,\label{intex}  \ee the first term displays the ultraviolet divergence in the limit $\widetilde{\epsilon}\rightarrow 0$, therefore we choose the counterterm $\Delta^{(1)}_m = -\ln(\widetilde{\epsilon})$ to cancel it. In the second term we have explicitly taken the limit $\widetilde{\epsilon}\rightarrow 0$ because the logarithmic singularity as $x'\rightarrow x$ is integrable. After renormalization, the equation (\ref{firste}) for $\X^{(1)}(x)$ becomes
\be \Big[\frac{d^2}{dx^2} + \frac{x^2}{4} \Big]\X^{(1)}(x)    =  S(x)~~;~~ S(x)=  x\,\int^x_0 \ln[x-x'] \,\frac{d}{dx'}\Big[ x'\,\X^{(0)}(x')\Big]\,dx' \,.\label{renX1eq}  \ee It is convenient to change integration variables, defining $x' \equiv x z$, and use the definition (\ref{chizi}) yielding
\be S(x) =  \widetilde{A}\,x^3\,\ln(x) V(x) +\widetilde{A}\,x^3\,\int^1_0 \ln[1-z] \,\frac{d}{dz}\,\Big[z^2\,V(zx)\Big]\,dz  \,,\label{Sfin}\ee for $x\ll 1$ the leading order contribution is from the first term featuring $\ln(x)$, since the second term features only powers of $x$ with $V(zx) \simeq 1 - \frac{z^4\,x^4}{80} +\cdots$ and the integrals do not yield large logarithms in the super-Hubble limit. The first term in eqn. (\ref{Sfin}) is the secular term, therefore it follows that to leading order in the super-Hubble limit
\be S(x) =  \widetilde{A}\,x^3\,\ln(x)  \,.\label{slead}\ee

The solution of the inhomogeneous equation (\ref{renX1eq}) is given by
\be \X^{(1)}(x) = \int^\infty_0 G_R(x,x')\,S(x')dx'\,, \label{soluX1}\ee where $G_R(x,x')$ is the retarded Green's function of the differential operator $\frac{d^2}{dx^2} + \frac{x^2}{4}$. It is constructed from the two linearly independent solutions of the homogeneous equation   (\ref{yeven},\ref{yodd}), and  given by
\be G_R(x,x')= \Big[ y_o(x)\,y_e(x')-y_e(x)\,y_o(x')\Big]\,\Theta(x-x')\,,\label{Gret}\ee therefore, we find
\be \X^{(1)}(x)= \widetilde{A} \Bigg\{y_o(x)\,\int^x_0 {x'}^3 y_e(x')\,\ln(x')\,dx' - y_e(x)\int^x_0 {x'}^3\,y_o(x')\,\ln(x')\,dx' \Bigg\}\,. \label{gsol}\ee Again, it is convenient to rescale the integration variable $x'=z x$ and to leading order in $x$ and $\ln(x)$ we find
\be \X^{(1)}(x)= \frac{\widetilde{A}}{20}\, x^5 \ln(x) \,\Big[1+ \mathcal{O}(1/\ln(x))+\cdots\Big]  \,.\label{leadchi1}\ee The $\ln(x)$ is yields secular growth as compared to a similar power in the unperturbed solution.

\textbf{Finite temperature contribution:}
The origin of the logarithmic enhancement in the source term (\ref{slead}) is the short distance behavior of the self-energy as $x' \rightarrow x$ (or $\eta' \rightarrow \eta$). In the super-Hubble limit  $x,x' \rightarrow 0$, the behaviour of the finite temperature contribution is given by eqn. (\ref{ITfin}) and it can be expanded in odd powers of the argument. Therefore, when input into the source term, it is straightforward to show that the leading contribution from the finite temperature correction is of $\mathcal{O}(x^5)$ without large logarithms for $x\ll1$, hence subleading with respect to  (\ref{slead}).

Therefore to leading order in $x$ and $\ln(x)$ we find in the super Hubble limit ($x\ll 1$)
\be \X(x) = \widetilde{A} x\Big[1-\frac{\widetilde{g}^2}{20}\,x^4\,\ln\big(\frac{1}{x} \big) + \cdots \Big] \,,\label{Xofxlead}\ee which suggests to absorb the secular term into a renormalization of the amplitude.

The dynamical renormalization group implements this as follows\cite{drggold,drg1,drg} (see appendix (\ref{app:drg})): write the amplitude $\widetilde{A}$ as
\be \widetilde{A} \equiv \widetilde{\mathcal{A}}[\overline{x}]\,Z[\overline{x}] ~~;~~ Z[\overline{x}]= 1+ \widetilde{g}^2\,z_1[\overline{x}]+ \cdots \,,\label{renA}\ee where $\overline{x}$ is an arbitrary renormalization scale, therefore
\be \X(x) = \widetilde{\mathcal{A}}[\overline{x}]\, x\, \Big[1+ {\widetilde{g}^2}\Big(z_1[\overline{x}]-\frac{1}{20}\,x^4\,\ln\big(\frac{1}{x} \big)\Big) + \cdots \Big] \,.\label{Xofxlren}\ee
The perturbative expansion is improved by choosing
\be z_1[\overline{x}]=\frac{1}{20}\,\overline{x}^4\,\ln\big(\frac{1}{\overline{x}} \big)\,,\label{z1}\ee which absorbs the secular term at the scale $\overline{x}$ and improves the perturbative expansion for  $\overline{x}$ close to $x$.

Since $\X(x)$ is independent of the renormalization scale $\overline{x}$ it obeys the dynamical renormalization group equation\cite{drg}
\be \frac{d}{d\overline{x}}\X(x) = 0 \,,\label{drgeq}\ee yielding

\be  \frac{d}{d\overline{x}}\widetilde{\mathcal{A}}[\overline{x}]\,x\, \Big[1+ {\widetilde{g}^2}\Big(z_1[\overline{x}]-\frac{1}{20}\,x^4\,\ln\big(\frac{1}{x} \big)\Big) + \cdots \Big]= -\widetilde{g}^2\,\widetilde{\mathcal{A}}[\overline{x}]\,x\, \frac{d}{d\overline{x}}z_1[\overline{x}] \,,\label{drgeq2}\ee which up to order $\mathcal{O}(\widetilde{g}^2)$ becomes
\be  \frac{d}{d\overline{x}}\widetilde{\mathcal{A}}[\overline{x}] = -{\widetilde{g}}^{\,2}\,\widetilde{\mathcal{A}}[\overline{x}]\, \frac{d}{d\overline{x}}z_1[\overline{x}]\,,\label{drgfineq}\ee with solution
\be \widetilde{\mathcal{A}}[\overline{x}]= \widetilde{\mathcal{A}}[0]\,e^{-\frac{{\widetilde{g}}^{\,2}}{20}\,\overline{x}^4\,\ln\big(\frac{1}{\overline{x}} \big)}\,.\label{renamp}\ee Finally, since $\overline{x}$ is arbitrary, we can now choose $\overline{x} = x$ yielding the dynamical renormalization group improved solution in the super Hubble limit
\be \X(x) = \widetilde{\mathcal{A}}[0]\,e^{-\frac{{\widetilde{g}}^{\,2}}{20}\, {x}^4\,\ln\big(\frac{1}{ {x}} \big)}\,x\Big[1 +\cdots \Big]  \,,\label{Xsoldrg}\ee where the dots stand for high powers of $x$.

In this limit, where the amplitude for the original unscaled field $\phi_1$ is constant in absence of coupling to the massless fields, using the relation (\ref{x2mt})  it follows   that this amplitude decays as
\be \phi_1(t) \propto \frac{\X(x)}{x} \propto e^{-\frac{g^2}{10} t^2\ln(1/m t)} \,.\label{suphubfi}\ee This result has a greater impact in the case of an ultralight particle, such as an axion-like dark matter candidate,  because in this case the time scale during which the Compton wavelength is larger than the Hubble radius is much longer leading to a larger suppression of the initial amplitude as a consequence of the decay prior to the beginning of oscillations. The coupling to massless (conformal) fields studied here is a \emph{proxy} for the case of an axion coupled to electromagnetic fields (which is conformally coupled in any cosmology), albeit with caveats discussed in section (\ref{sec:conclusions}).

We have argued above that a simple, local friction term cannot reliably describe the self-energy contribution to the equation of motion of the condensate, therefore the evolution of the condensate with a friction term is    quantitatively incorrect. We can now assess whether such phenomenological description is at least qualitatively reliable. With this purpose we re-write the equation (\ref{xvar}) in terms of the dimensionless variable $x$ related to conformal time $\eta$ via equation (\ref{xvar}). In terms of the renormalized mass $m_R$, we find \be \frac{d^2}{dx^2}\,\X(x) + \frac{x^2}{4}\,\X(x) + \Big(\frac{\Gamma}{2m_R} \Big) \, x^2 \,\frac{d}{dx}\Big( \frac{\X(x)}{x}\Big) =0 \,,\label{fricx} \ee where the zero temperature decay rate of the massive into two massless particles  is (see equation (\ref{masswidth}) in appendix \ref{app:drg}) $\Gamma = \pi g^2/2m_R$ (in terms of the renormalized mass) therefore, in terms of the dimensionless cooupling introduced in eqn. (\ref{nudimvars}) it follows that
\be \frac{\Gamma}{2m_R}  = \pi \widetilde{g}^2\,,\label{equifric}\ee and the source term on the right hand side of  equation (\ref{firste}) becomes
\be S_f(x) = -\pi  x^2 \,\frac{d}{dx}\Big( \frac{\X^{(0)}(x)}{x}\Big)  \,.\label{sourcefriction}\ee

 We follow the steps above and treat the friction term in perturbation theory, obtaining the set of equations (\ref{zerote},\ref{firste}) but now the source term from friction is to leading order in $x$
\be S_f(x)=  A\,\frac{\pi}{20}\,x^5  \,, \label{sourcefric}\ee  yielding
\be \X^{(1)}(x) \propto A\,x^7 \,,\label{fricX1}\ee in the super-Hubble limit, which obviously is not even qualitatively close to the result (\ref{slead}) which yields a much stronger relaxation. Finite temperature does not help, the finite temperature decay rate in Minkowski space time is (see appendix)
\be \Gamma(T) =   \frac{\pi\,g^2}{2m_R} \,\Big[1+2 n\Big(\frac{m_R}{2}\Big)\Big] \,,\label{GamaTm}\ee therefore the   source term (\ref{sourcefric}) is only modified by a multiplicative factor but the power of $x$ remains the same.  We conclude that a friction term does not reliably describe the decay of the condensate on super-Hubble scales.

This failure is not unexpected, even in Minkowski space-time, a decay rate is extracted in the long time (formally infinite time) limit. This is explicit from the textbook
S-matrix calculation which takes the infinite time limit yielding strict energy conservation, or at a more elementary level, from Fermi's Golden rule which takes the long
time limit to replace a strongly peaked function by an exact delta function. These long time limits completely neglect transient phenomena, that while may not be observationally relevant in Minkowski space-time,   are important in an expanding cosmology.

\vspace{1mm}
\subsection{Sub-Hubble limit, zero temperature:}\label{sub:sub}  We now focus on the sub-Hubble limit where $x \gg 1$ and study first the simpler case of zero temperature
 to establish a direct comparison with Minkowski space-time and with the phenomenological decay dynamics described by the friction term. In this limit it is more convenient to use the mode functions $U(x)$ given by eqn. (\ref{Uwkb}), with Wronskian $\mathcal{W}_x[U(x),U^*(x)] = -i$. The main reason for using these mode functions is that these allow a direct comparison with the Minkowski space-time case (see appendices).

We now must solve equation (\ref{firste}) with the source term (\ref{source}) and for the sub-Hubble case
\be \X^{(0)}(x) = A\, U(x) + A^* \,U^*(x) \,,\label{X0sub}\ee with
\be U(x) = \frac{e^{-i\frac{x^2}{4}}}{\sqrt{x}} \,, \label{Ufa}\ee and
$A$ a complex amplitude.

\vspace{1mm}
  The zero temperature contribution to the self energy is given by the first term on the right hand side of eqn. (\ref{tilsig}). It is convenient to change variables to $\tau \equiv x-x'$ yielding
\be\int^x_0 \widetilde{\Sigma}_0(x,x')\,U(x')\,dx'  = -x^2\,U(x)\,\int^x_0\Bigg\{\frac{\tau}{\tau^2+\widetilde{\epsilon}^2} + \frac{\tau}{\tau^2+\widetilde{\epsilon}^2} \Bigg[ \sqrt{1-\frac{\tau}{x}}\, \Big(e^{i\frac{x}{2}\tau\big(1-\frac{\tau}{2x}\big)} -1 \Big) \Bigg] \Bigg\}\,d\tau \,, \label{SigU}\ee where $\widetilde{\Sigma}_0$ is the zero temperature contribution of the total self-energy (\ref{tilsig}), the contribution from the term with $U^*$ in $\X^{(0)}$ is simply the complex conjugate.

The first term inside the integral yields
in the limit $\widetilde{\epsilon}\rightarrow 0$
\be \int^x_0 \frac{\tau}{\tau^2+\widetilde{\epsilon}^2}\, d\tau = \frac{1}{2}\,\ln\big(\frac{x^2}{2}\big) - \frac{1}{2}\,\ln\big(\frac{\widetilde{\epsilon}^2}{2}\big)\,,\label{divsub}\ee again the ultraviolet divergence of the second term in this integral is cancelled by a proper choice of the counterterm $\Delta^{(1)}_m$ in the source (\ref{source}), in the second term  in (\ref{SigU}) we can safely take $\widetilde{\epsilon}^2 \rightarrow 0$ because there are no divergences as $\tau \rightarrow 0$.

After renormalization we find
\be S(x) = A\,U(x)\,\mathcal{J}[x] + A^*\,U^*(x)\,\mathcal{J}^*[x]\,,\label{Ssub}\ee where upon changing integration variables  $ z =  \frac{x \tau}{2} $ we obtain
\be \mathcal{J}[x] = x^2\,\Bigg\{\frac{1}{2}\,\ln\big(\frac{x^2}{2}\big) + \int^{\frac{x^2}{2}}_0\,\Big\{\sqrt{1-\frac{2 z}{x^2}}\,\cos\Big[z \big(1-\frac{z}{x^2} \big)  \Big]-1 \Big\} \frac{dz}{z} \, + i   \int^{\frac{x^2}{2}}_0\, \sqrt{1-\frac{2 z}{x^2}}\,\sin\Big[z \big(1-\frac{z}{x^2} \big)  \Big]  \frac{dz}{z}\Bigg\}\,.\label{bigJ} \ee At this stage we can compare these results with those in the case of Minkowski space-time in appendix (\ref{app:drg}), in particular comparing $\mathcal{J}(x)$ eqn. (\ref{bigJ}) with $\mathcal{F}(t)$ equation (\ref{foftex2}) with the identification  $\frac{x^2}{4} = mt$ (see equation (\ref{x2mt})). In the formal limit $x^2 \rightarrow \infty$ the cosine and sine terms in eqn. (\ref{bigJ}) become identical to the cosine and sine terms of in equation (\ref{foftex2}), this is the infinite time limit where these terms yield secular contributions that require resummation via the dynamical renormalization group. The prefactor $x^2$, the square roots inside the integrands and the terms $z/x^2$ in the arguments of the cosine and sine functions in eqn. (\ref{bigJ}) can all be traced back to the expansion scale factors.

In order to obtain $\X^{(1)}(x)$ from equation (\ref{soluX1}) we need the retarded Green's function, which is now constructed from the solutions $U(x),U^*(x)$ of the homogeneous equation (\ref{zerote}) and is given by
\be G_R(x,x')= i\Big[U(x)\,U^*(x')-U^*(x)\,U(x') \Big]\,\Theta(x-x') \,,\label{Gret2}\ee  from which we obtain
\bea \X^{(1)}(x) & = &   U(x)\,i\int^x_0 U^*(x')\,\Big[A\,U(x')\,\mathcal{J}[x']+A^*\,U^*(x')\,\mathcal{J}^*[x']\Big]\,dx'\nonumber \\
& - &  U^*(x)\,i\int^x_0 U(x')\,\Big[A\,U(x')\,\mathcal{J}[x']+A^*\,U^*(x')\,\mathcal{J}^*[x']\Big]\,dx'\,.\label{X1sub}\eea We seek to identify the secular terms in the limit $x\rightarrow \infty$. With the sub-Hubble mode functions (\ref{Ufa}), it follows that
\be U^*(x')U(x') = \frac{1}{x'}~~;~~ U(x')U(x') = \frac{e^{-i\frac{x^{'\,2}}{2}}}{x'} \,,\label{prods}\ee therefore, just as in the case of Minkowski space time (see appendix (\ref{app:drg})), the second type of products feature rapid oscillations in the long time limit,   do not yield secularly growing terms and remain perturbative at all time, whereas   the products $U^*(x') U(x')$ do not feature these rapid oscillations and yield the secular contributions. Isolating the secular terms, we find
\be \X^{(1)}_s(x) = A\,U(x)\Big[i\int^x_0  \,\frac{\mathcal{J}[x']}{x'}\,dx'\Big]  + c.c. \,,\label{X1sec}\ee yielding
\be \X(x) = A\,U(x)\,\Big[1 + i\widetilde{g}^2\int^x_0  \,\frac{\mathcal{J}[x']}{x'}\,dx'\Big] + A^*\,U^*(x) \Big[1 - i\widetilde{g}^2\int^x_0  \,\frac{\mathcal{J}^*[x']}{x'}\,dx'\Big] + \mathrm{non~secular} \,.\label{Xsubfin}\ee Proceeding as in the previous case
we write
\be A = \mathcal{A}[\overline{x})Z[\overline{x}]~~;~~Z[\overline{x}]= 1 + \widetilde{g}^2\,z_1[\overline{x}] +\cdots \,\label{amprensub}\ee and choosing $z_1[\overline{x}]$ to cancel the integrals yielding secular terms at the renormalization scale $\overline{x}$ leads to the improved expression

\be \X(x) = \mathcal{A}[\overline{x}]\,U(x)\,\Big[1 + i\widetilde{g}^2\int^x_{\overline{x}}  \,\frac{\mathcal{J}[x']}{x'}\,dx'\Big] + \mathcal{A}^*[\overline{x}]\,U^*(x) \Big[1 - i\widetilde{g}^2\int^x_{\overline{x}}  \,\frac{\mathcal{J}^*[x']}{x'}\,dx'\Big] + \mathrm{non~secular}\,.\label{Srensub}\ee The dynamical renormalization group equation (\ref{drgeq}) up to $\mathcal{O}(\widetilde{g}^2)$ now leads to the amplitude equation
\be \frac{d}{d\overline{x}}\mathcal{A}[\overline{x}] = i\,\widetilde{g}^2\,\mathcal{A}[\overline{x}]\, \frac{\mathcal{J}[\overline{x}]}{\overline{x}}\,, \label{drgAsub}\ee with solution
\be \mathcal{A}[\overline{x}]=\mathcal{A}[\overline{x}_o]\,e^{i\,\widetilde{g}^2\int^{\overline{x}}_{\overline{x}_o}\frac{\mathcal{J}[x']}{x'}\,dx' } \,.\label{ampsolsub} \ee Choosing the arbitrary scales $\overline{x}\equiv x;\overline{x}_o\equiv x_o$ we find the improved solution
\be \X(x) = \mathcal{A}[ {x}_o]\,U(x)\,e^{i\,\widetilde{g}^2\int^{ {x}}_{ {x}_o}\frac{\mathcal{J}[x']}{x'}\,dx' }+\mathrm{c.c.} \,.  \label{Xrensub}\ee We are primarily interested in the decay of the amplitude of the condensate, which is determined by
\be e^{-\frac{\gamma(t)}{2}} \equiv e^{-\frac{\widetilde{g}^2}{2}\int^{ {x^2(t)}}_{ {x}^2_o}\frac{ \mathcal{J}_I[x']}{{x'}^2}\,d{x'}^2 }\,,\label{gafu}\ee  where $x^2(t)= {4mt}$ (see equation (\ref{x2mt})),  and $\mathcal{J}_I[x]$ is the imaginary part of $\mathcal{J}$, namely
 \be \frac{ \mathcal{J}_I[x']}{{x'}^2}=  \int^{\frac{x^{'\,2}}{2}}_0\, \sqrt{1-\frac{2 z}{x^{'\,2}}}\,\sin\Big[z \big(1-\frac{z}{x^{'\,2}} \big)  \Big]  \frac{dz}{z}\,.\label{JIx}\ee

 Note the similarity of the integral to the case of Minkowski space time in appendix (\ref{app:drg}), again, the factors $z/x'^2$ are a consequence of the scale factor and reflect the effect of the cosmological expansion.

 We define the \emph{time dependent} decay rate as
 \be \Gamma(t) \equiv \frac{d}{dt}\gamma(t) = \frac{g^2}{2m}\, \Bigg(2 \,\frac{ \mathcal{J}_I[x]}{{x}^2}\Bigg)\Big|_{x^2=4mt} \,,\label{rateoft}\ee we refer to $\gamma(t)$ as the \emph{decay function} and $\frac{ \mathcal{J}_I[x]}{{x}^2}$ as the \emph{decay rate function}.

\begin{figure}[ht!]
\begin{center}
\includegraphics[height=3in,width=4in,keepaspectratio=true]{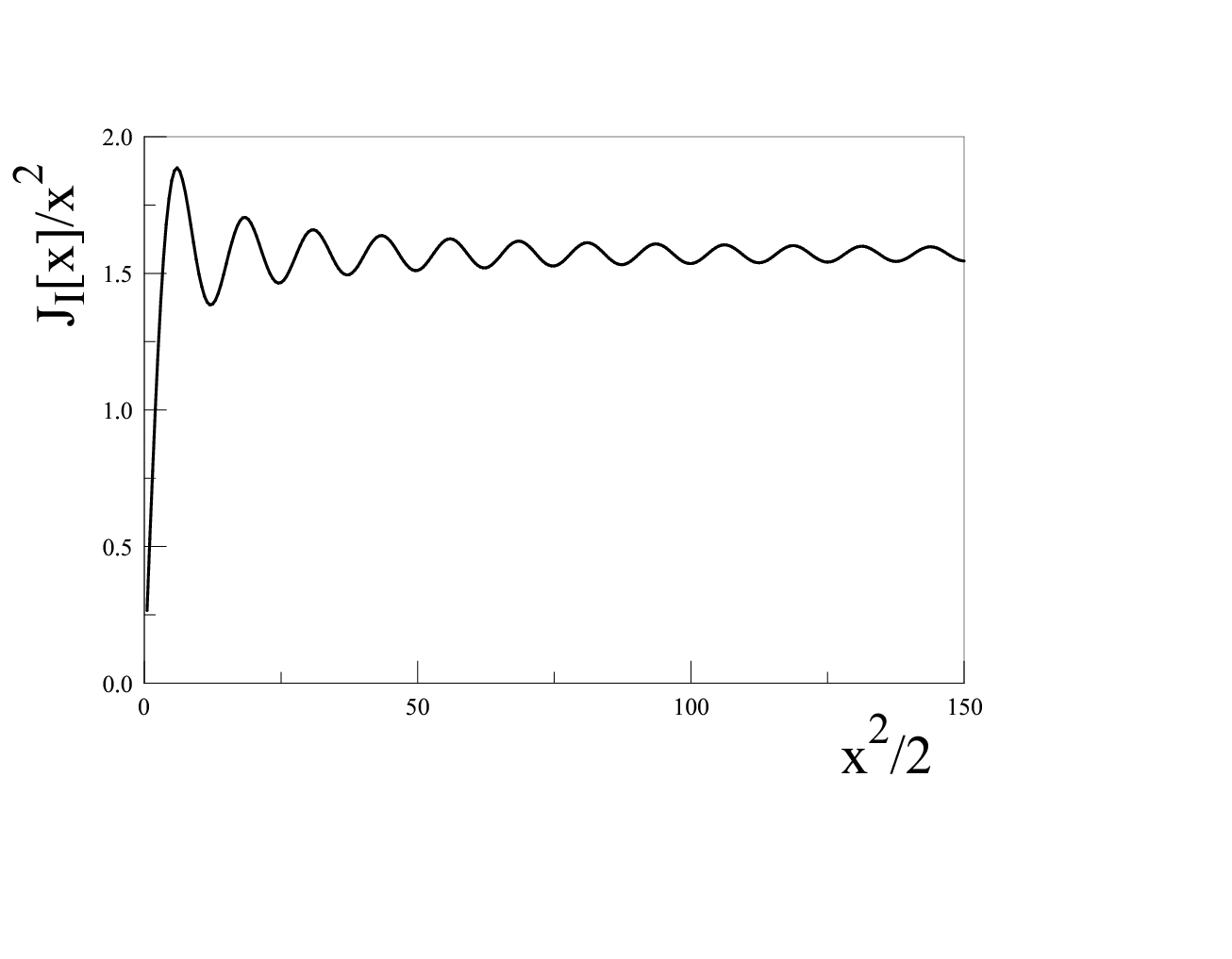}
\caption{ The integral $J_I[x]/x^2$ in eqn. (\ref{inftylim}) vs. $x^2/2$. The asymptotic limit is $\pi/2$ . }
\label{fig:sineint}
\end{center}
\end{figure}

 In the formal limit $x\rightarrow \infty$ we find
 \be \frac{ \mathcal{J}_I[x]}{{x^2}}=\int^{\frac{x^2}{2}}_0\, \sqrt{1-\frac{2 z}{x^2}}\,\sin\Big[z \big(1-\frac{z}{x^2} \big)  \Big]  \frac{dz}{z} {~~~\overrightarrow{x  \rightarrow \infty}}~~~ \frac{\pi}{2}\,, \label{inftylim}\ee yielding
 \be \Gamma(t)_{~~\overrightarrow{t  \rightarrow \infty}}~~ \frac{\pi\,g^2}{2m}\,,\label{asyrateoft}\ee  which is precisely the zero temperature decay rate in the Minkowski space-time limit, see equations (\ref{reimpole}, \ref{masswidth}) for  zero temperature in the appendices.  The decay rate function $\frac{\mathcal{J}_I[x]}{x^2} $
  is displayed in fig.(\ref{fig:sineint}), exhibiting the asymptotic behavior (\ref{inftylim}).

 Therefore in the $x \rightarrow \infty$ limit we would expect
  \be \gamma(t) _{~~\overrightarrow{mt \rightarrow \infty}}~~  \Gamma  t ~~;~~ \Gamma =   \frac{\pi g^2}{2m}\,, \label{minli}\ee recognizing the zero temperature exponential decay law for the amplitude with the decay rate from Minkowski space time. This result is expected since the infinite time limit corresponds to the deep sub-Hubble regime where the mode functions are very similar to Minkowski space-time. However, we are interested in understanding transient phenomena describing the decay of the amplitude in the sub-Hubble regime but still at finite time, and to compare this behavior to the Minkowski space-time limit.

The equation of motion with the phenomenological friction term is given by eqn. (\ref{fricx})  with the equivalence (\ref{equifric}), yielding the source term (\ref{sourcefriction}) for equation (\ref{firste}). For the sub-Hubble mode functions (\ref{Ufa}) and $\X^{(0)}(x)$ given by eqn. (\ref{X0sub}) we find that the source term $S(x)$ is of the same form as in eqn. (\ref{Ssub}) but now with
\be \mathcal{J}_f[x] = \frac{\pi}{2} \Big[ 3+  i  {x^2}     \Big] \,,\label{fricJ}\ee from which we obtain the decay function
\be  \frac{\mathcal{J}_{fI}[x]}{{x}^2}= \frac{\pi}{2}\,  \,,\label{ImJf}\ee which is precisely the asymptotic limit for large $x'$ of equation (\ref{JIx}) as follows from (\ref{inftylim}). Inserting the result  (\ref{ImJf}) into the decay function $\gamma(t)$ defined in eqn. (\ref{gafu}) we find the decay function arising from the friction term (with $x_o=0$)  \be \gamma_f(t) = \pi \widetilde{g}^2\, (2mt)\equiv \Gamma t \,,\label{gamf}\ee  where we used the relation (\ref{equifric}). This result is not surprising, the sub-Hubble modes to leading order in the (WKB) expansion are the closest to Minkowski space time, and the friction term is given by the decay rate obtained in the infinite time limit. Figure (\ref{fig:sineint}) with $x^2/2=mt$ clearly shows that the rate approaches the Minkowski space time limit at a time scale $t \gtrsim 100/m$, which for an ultralight dark matter candidate may be very long, with the conclusion that transient dynamics during the expansion is important during long time scales for ultralight particles.

An important aspect of the result (\ref{JIx})   is that the square root factors lead to a smaller decay rate function and decay function, consequently a \emph{slower relaxation} as compared to its Minkowski space-time limit given by the decay function $\mathcal{F}(t)$ in equation (\ref{foftex2}). Therefore, whereas the relaxation function $\mathcal{J}$ tends asymptotically to the Minkowski limit, during cosmological expansion, the relaxation and decay are \emph{slower} than in the Minkowski space-time limit.  Therefore a main conclusion is that  the local friction term \emph{overestimates} the decay rate and underestimates the lifetime of the condensate.

\begin{figure}[ht!]
\begin{center}
\includegraphics[height=3in,width=4in,keepaspectratio=true]{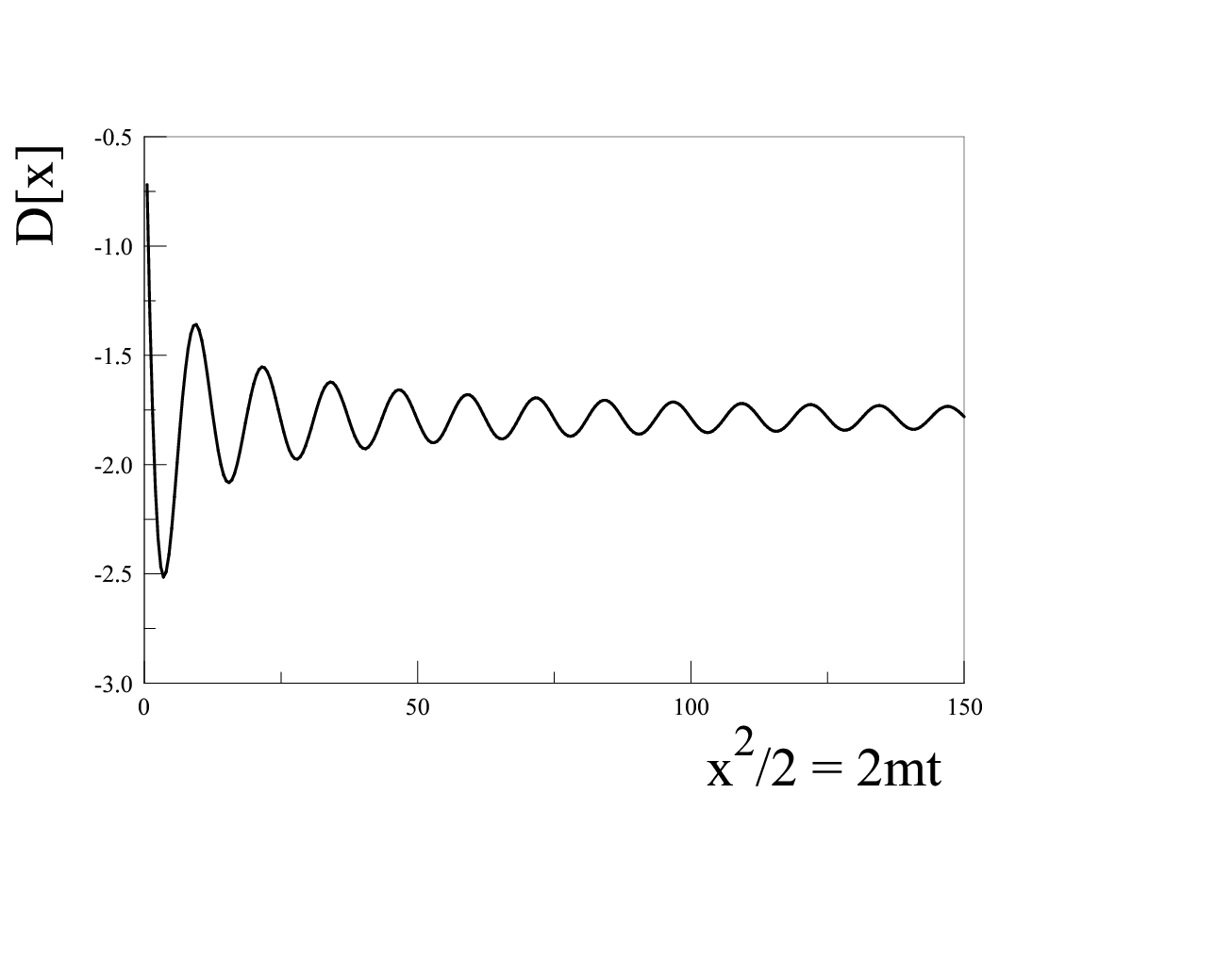}
\caption{ The function $D[x]$ defined by eqn. (\ref{Dofx}) vs. $x^2/2=2mt$.  }
\label{fig:diference}
\end{center}
\end{figure}

In order to understand the \emph{transient dynamics}, and to quantify the difference between the decay functions with and without cosmological expansion,  we now consider the \emph{difference} between the decay rate functions which are determined by the integral
\be  {D}[x] = \frac{1}{2}\int^{x^2}_0 \Big[ \frac{\mathcal{J}_{I}[x']}{{x'}^2}  -\frac{\mathcal{J}_{fI}[x']}{{x'}^2} \Big] \,d{x'}^2 \,,\label{Dofx}\ee this function is displayed in fig. (\ref{fig:diference}) showing that the decay function from the zero temperature friction term always \emph{overestimates} the decay of the amplitude. This is one of the important results from this study.

The difference $D[x]$ displayed in fig. (\ref{fig:diference}) also highlights that at long time $t \gg 1/m$ the ratio of the difference to the decay function $D[t]/\gamma_f(t) \propto 1/(mt)$. Therefore in the (very) long time limit $t \gg 1/m$,   the decay function including the cosmological expansion approaches asymptotically the decay function from the simple friction term but always from below. This is of course expected because   the (very) long time limit corresponds to the deep sub-Hubble limit wherein the equivalence principle dictates that the dynamics should be insensitive to the Hubble radius. However, at early and intermediate times $mt  \lesssim 100$ there is a substantial departure from the decay law obtained from the phenomenological friction term, a consequence of the cosmological expansion. Hence, for ultralight dark matter, the time scales where transient dynamics is very important and cannot be described by a local friction term may indeed be very long.

\vspace{1mm}

\subsection{Sub-Hubble, finite temperature:}
\begin{figure}[ht!]
\begin{center}
\includegraphics[height=3in,width=3in,keepaspectratio=true]{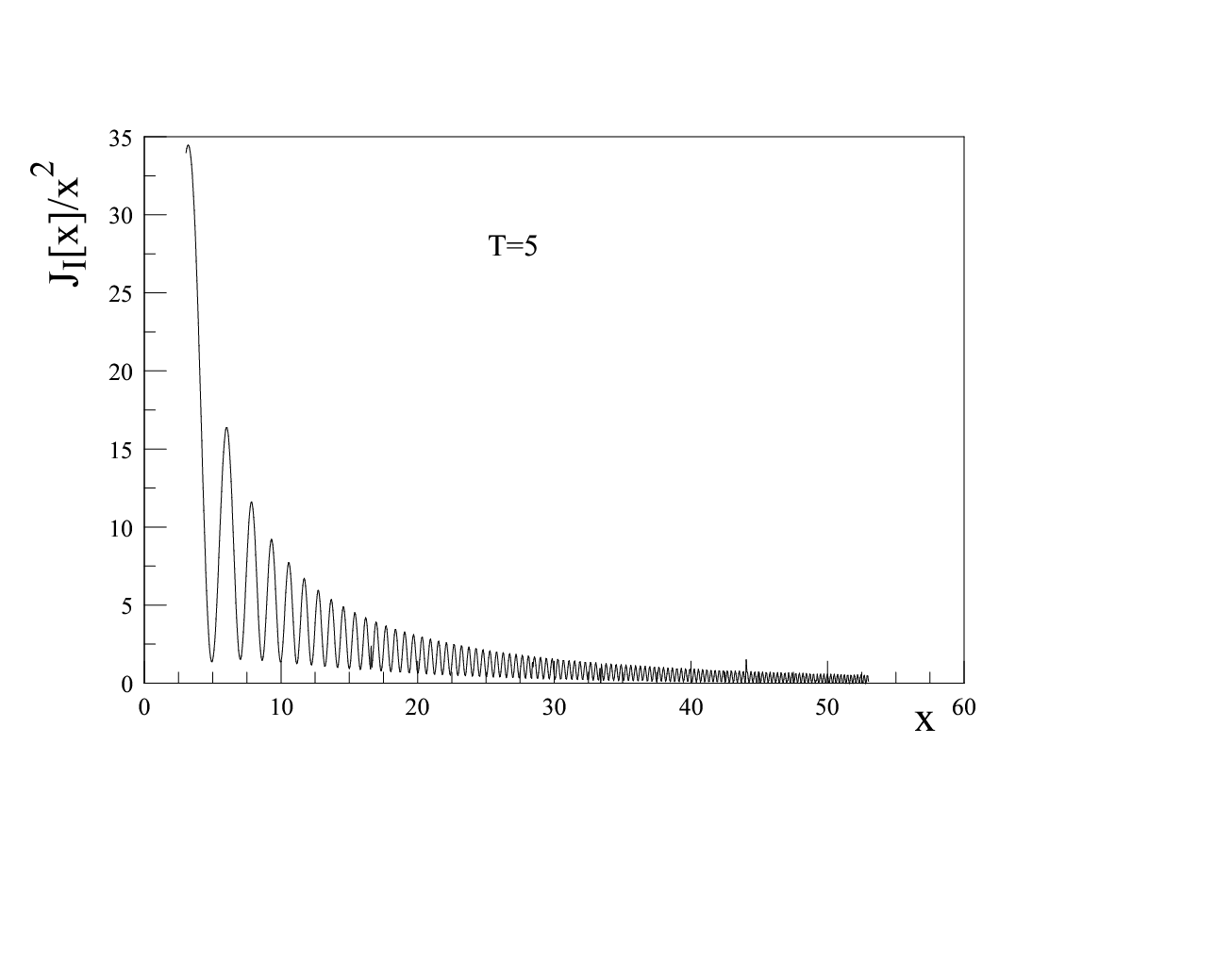}
\includegraphics[height=3in,width=3in,keepaspectratio=true]{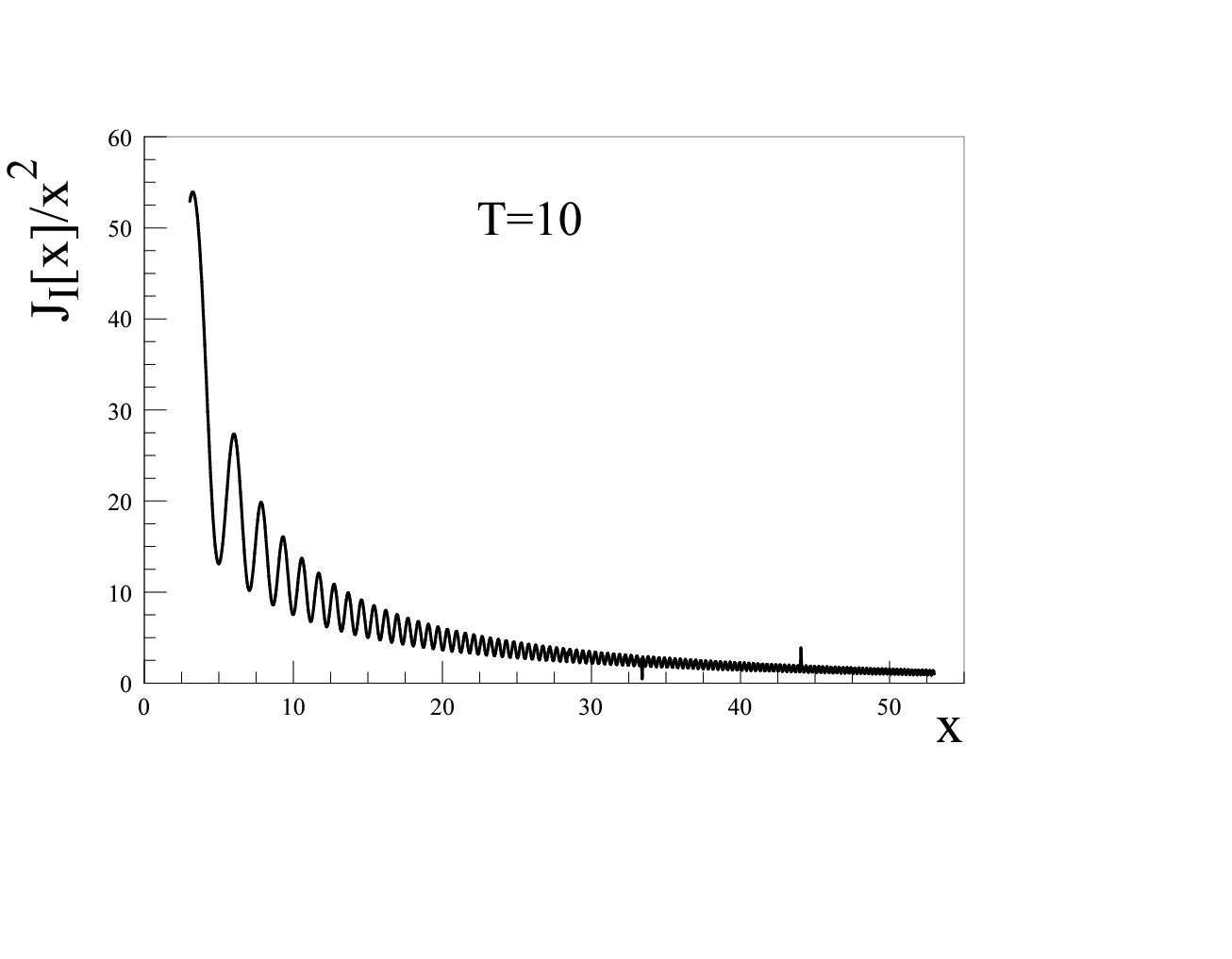}
\caption{ The finite temperature contribution to the  rate function $\mathcal{J}_I[x]/x^2$ defined by eqn. (\ref{rateJT}) vs. $x$ for $\widetilde{T}= 5,10$.  }
\label{fig:JI}
\end{center}
\end{figure}

The finite temperature contribution to the self energy is given by the second term inside the brackets in eqn. (\ref{tilsig}), from which, defining $x-x'=\tau$, we find
\be \int^x_0 \widetilde{\Sigma}_T(x,x')\,U(x')\,dx'  = -x^2\,U(x)\,(2\pi \widetilde{T})\int^x_0\Bigg\{\Bigg[ \coth\big[2\pi \widetilde{T} \tau\big]   - \frac{1}{2\pi \widetilde{T}\tau}\Bigg]\,\sqrt{1-\frac{\tau}{x}}\,\,  e^{i\frac{x}{2}\tau\big(1-\frac{\tau}{2x}\big)}     \Bigg\}\,d\tau \,, \label{SigUT}\ee which yields a source term of the same form as (\ref{Ssub}) with $\mathcal{J}[x]$ now given by
\be \mathcal{J}[x] = x^2 \, \Big(4\pi \frac{\widetilde{T}}{x}\Big)\int^{\frac{x^2}{2}}_0\Bigg\{\Bigg[ \coth\big[4\pi \frac{\widetilde{T}}{x} z\big]   - \frac{1}{4\pi \frac{\widetilde{T}}{x}z}\Bigg]\,\sqrt{1-2\frac{z}{x^2}}\,\,  e^{iz\big(1-\frac{z}{ x^2}\big)}     \Bigg\}\,dz\,,\label{JT} \ee after changing variable to $z= x\tau/2$. The rate function becomes
\be \frac{\mathcal{J}_I[x]}{x^2} =  \Big(4\pi \frac{\widetilde{T}}{x}\Big)\int^{\frac{x^2}{2}}_0\Bigg\{\Bigg[ \coth\big[4\pi \frac{\widetilde{T}}{x} z\big]   - \frac{1}{4\pi \frac{\widetilde{T}}{x}z}\Bigg]\,\sqrt{1-2\frac{z}{x^2}}\,\, \sin\Big[z\big(1-\frac{z}{ x^2}\big)  \Big]      \Bigg\}\,dz\,.\label{rateJT} \ee  A remarkable aspect of this expression is that it depends on temperature via the ratio
\be \frac{\widetilde{T}}{x} = \frac{T(\eta)}{2m}~~;~~ T(\eta) = \frac{T}{  a(\eta)}\,,\label{Tofeta}\ee where we used the definitions (\ref{nux},\ref{nudimvars}),  namely the temperature emerges redshifted by the scale factor. This is important, in the long time limit $x\rightarrow \infty$ the effective temperature $T(\eta) \rightarrow 0$ and the finite temperature contribution to the decay rate effectively vanishes. This phenomenon is displayed in fig. (\ref{fig:JI}).

\begin{figure}[ht!]
\begin{center}
\includegraphics[height=3in,width=3in,keepaspectratio=true]{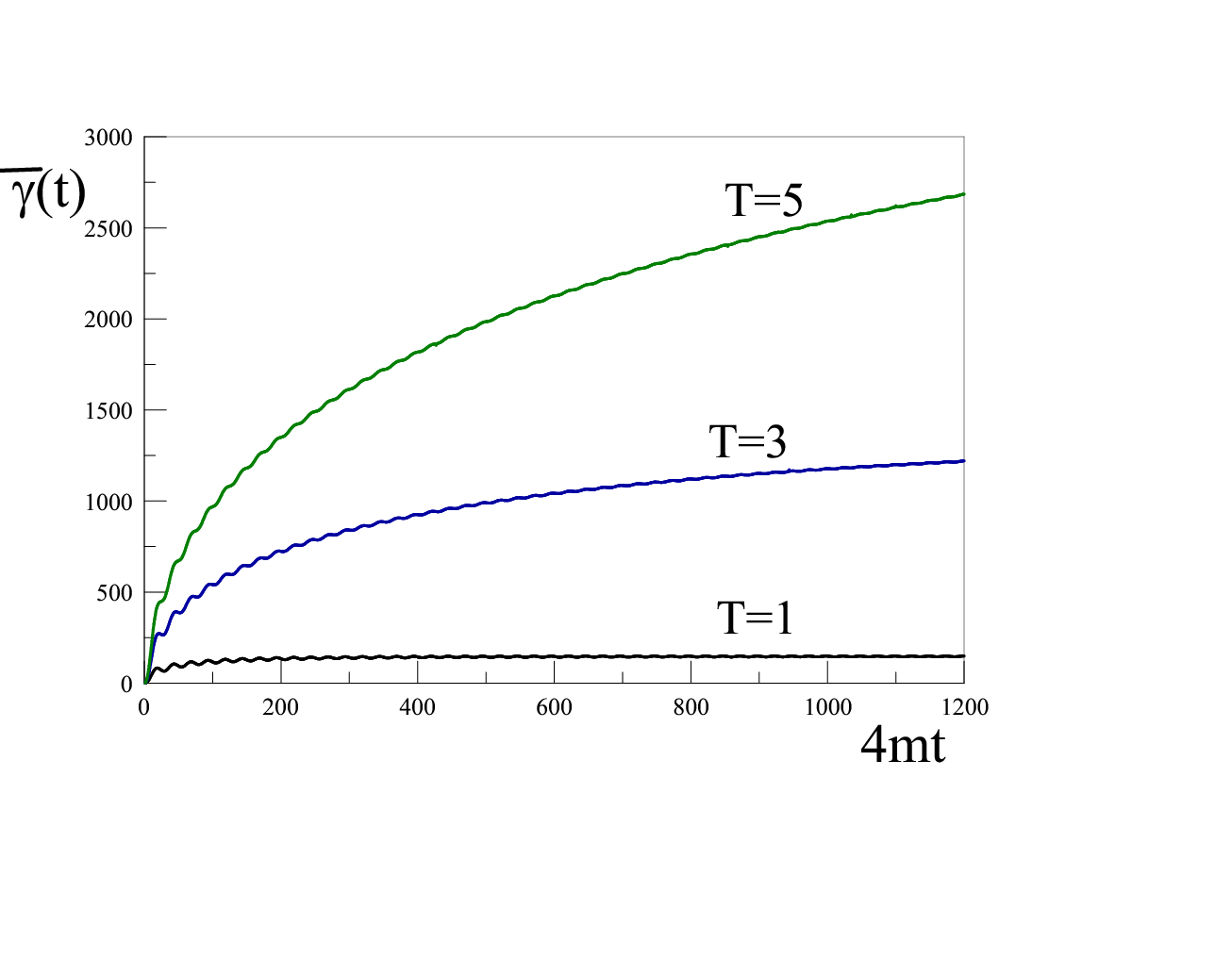}
\caption{ The finite temperature contribution to the decay function  $\overline{\gamma}(t) = \frac{\gamma(t)}{\widetilde{g}^2}=\int^{ {x^2(t)}}_{ 0}\frac{ \mathcal{J}_I[x']}{{x'}^2}\,d{x'}^2  $ (see equation(\ref{gafu})) vs. $x^2(t)=4mt$ for $\widetilde{T}=1,3,5$.  }
\label{fig:bargamma}
\end{center}
\end{figure}

These figures clearly show that the time dependent rate while increasing with $\widetilde{T}$, vanishes at long time, a behaviour with profound consequences for the decay function $\gamma(t)$ defined  by eqn. (\ref{gafu}): this function initially grows and eventually saturates reaching a plateau at a value that increases with $\widetilde{T}$ as displayed in fig. (\ref{fig:bargamma}). At long time this behavior yields a finite renormalization of the amplitude, but the secular growth of the zero temperature contribution dominates the  decay function at long time.

Taken together figures (\ref{fig:JI},\ref{fig:bargamma}) clearly show that by the time when the decay rate function approaches the Minkowski space-time limit $t \gtrsim 100/m$ the temperature has redshifted almost to zero. The enhancement of the decay function from  the  temperature contribution is manifest only during the transient dynamics which is not captured by the local friction term.  Hence, we conclude that the decay of the amplitude at long time $t \gtrsim 100/m$  is completely determined by the \emph{zero temperature limit} of the self-energy.

We highlight that the time dependence of the temperature $T(t) = T/a(t)$  in the self-energy correction is by no means of straightforward interpretation as the temperature expected to be  redshifted by the cosmological expansion:  the distribution function of the massless field, eqn. (\ref{nofk}) is time independent, therefore the temperature factors in the self-energy loop do \emph{not} depend on time, this is explicit in eqns. (\ref{sigi},\ref{IntI}) for the self-energy. Instead, the time dependence emerges from the \emph{convolution} of the self-energy and the unperturbed mode functions, eqn. (\ref{SigUT}),  whose time dependence ultimately feeds into the source term in the perturbed equations of motion as evidenced by the expressions (\ref{JT}).

A Minkowski space-time phenomenological friction term with a constant finite temperature and decay rate cannot reliably capture this behavior with rich and complex transient dynamics leading to a vanishing finite temperature contribution to the decay rate function at long time, thereby, again  \emph{overestimating} the decay rate. A phenomenological ``improvement'' of the local friction term would modify the temperature by the scale factor to account for the redshifting of the temperature with expansion.  However, this is an    \emph{a posteriori} and \emph{ad hoc} modification which is  incompatible with the calculation of the friction term in Minkowski space time, in equilibrium at finite temperature and via the S-matrix formulation which takes the infinite time limit.

\section{Discussion}\label{sec:discussion}

\textbf{Matching super to sub-Hubble:} As discussed above, the complexity of the mode functions prevent us from providing an analytic treatment of amplitude decay for all values of $x$, analyzing instead both the super-Hubble ($x\ll 1$) and sub-Hubble ($x\gg 1$) regimes separately, since the mode functions $U(x)$ are simpler in these regimes. A formal solution that interpolates between the super and sub-Hubble limits and therefore provides a matching between the two regimes is constructed as follows. Instead of the approximate forms valid in these limits, let us use the full complex solutions $U_0(x) \equiv U_{\kappa=0}(x)$ and its complex conjugate given by the $\kappa =0$ limit of equations (\ref{U},\ref{Ucc}) and write the unperturbed solution as in equation (\ref{X0sub}) with $U(x)$ replaced by $U_0(x)$. Requesting that in the  the $x \ll 1$ limit the unperturbed solution matches the super Hubble one given by (\ref{chizi}) fixes the amplitude $A$ in terms of $\widetilde{A}$. Secondly, cancel the divergence from the short distance expansion of the self-energy, namely the first term on the right hand side of eqn. (\ref{SigU}) with $U(x)$ replaced by $U_0(x)$ and   let us now \emph{define}
\be  {\mathcal{J}}[x] \equiv U^{-1}_0(x) \int^x_0 \widetilde{\Sigma}_0(x,x')\,U_0(x')\,dx' \,, \label{newjay}\ee so that
\be \int^x_0 \widetilde{\Sigma}_0(x,x')\,U_0(x')\,dx'  = U_0(x) \,{\mathcal{J}}[x] \,,\label{usig}\ee it is straightforward to show that this reproduces the source term both in the super and sub Hubble limits.
Now the source term is of  the same form as in eqn. (\ref{Ssub}) with $U(x)$ replaced by $U_0(x)$, the amplitude $A$ already fixed by the super-Hubble limit, and the first order perturbation $\X^{(1)}(x)$ features the same form as in eqn. (\ref{X1sub}) with the same replacement. The implementation of the dynamical renormalization group now follows the steps in section (\ref{sec:drg}).   This procedure yields the improved solution that differs  from the solution obtained in the previous section in the sub and super Hubble limits   by non-secular terms, namely always finite at all times  and which are subleading in both limits. While this formal procedure provides a systematic matching and describes the region $x\simeq 1$, the secularly growing terms in super and sub Hubble limits must be extracted by taking the corresponding limits of the mode functions, this is precisely what was achieved in the previous section. Therefore, although this matching framework interpolates between these limits, to characterize the   region $x\simeq 1$ the full expression of the (complicated) mode functions $U_0(x)$  must be taken into account. However, since our focus is to study the validity of the local friction term in both limits, the analysis in the previous section    provides a clear quantitative treatment and exhibits the shortcomings of the phenomenological approach in the super and sub-Hubbble limits in the simplest and clear manner.

\vspace{1mm}

\textbf{Scale factor dependence of the interaction:} The model studied in this article, defined by the Lagrangian density  (\ref{action}), is super renormalizable, the coupling $\lambda$ has mass dimension one. Passing to conformal time and conformally rescaling the fields, the interaction vertex becomes $\lambda a(\eta)$ as displayed by the conformal action (\ref{conformalaction}). If the coupling has mass dimension $\nu$, the corresponding conformal action features the coupling multiplied by $(a(\eta))^\nu$, for a renormalizable interaction $\nu=0$, but for a non-renormalizable one $\nu < 0$, implying that the effective interaction becomes stronger at earlier time. This is important for axions coupled to gauge fields via a Chern-Simons term $\propto \vec{E}\cdot\vec{B}$ with a non-renormalizable coupling of mass dimension $\nu =-1$, which may lead to a very different decay dynamics, and clearly merits further study.

\vspace{1mm}

\textbf{On axions:} We argued that the model studied in this article serves as  a \emph{proxy} for axions interacting with gauge fields. However, there are important caveats in the similarity that must be addressed before extrapolating the results to the case of axions. First, as discussed above, the interaction between axions and gauge fields via
a Chern-Simons term is non-renormalizable, leading to a very different dependence of the interaction   on the scale factor upon conformal rescaling. Secondly, we considered strictly massless fields in the self-energy loop, however gauge fields are screened by loops of charged leptons or quarks in the medium, therefore the gauge field contribution to the
self energy will be different from that of strictly massless fields at high temperature. Screening will become ineffective for temperatures well below $\simeq \mathrm{MeV}$ corresponding to electron-positron loops that screen the gauge fields. Therefore the similarity of the \emph{proxy} model to that of axions may be more clear and reliable during the radiation dominated era when $T \ll \mathrm{MeV}$ and screening of gauge fields is not effective.

\vspace{1mm}

\textbf{Impact on particle production and energy transfer:} The decay of the condensate amplitude is a consequence of the emission and recombination of massless quanta. This process implies the production of these quanta and an energy transfer from the condensate to the massless field. In fact the self-energy in the equation of motion describes the damping of the amplitude of the condensate as a consequence of the radiation reaction.   One of the main results obtained in this study is that as a consequence of the cosmological expansion  the decay function is \emph{smaller} than that obtained from the S-matrix approach in Minkowski space time (the local friction term). A direct corollary is that the transfer of energy from the condensate to the massless fields occurs more \emph{slowly} than would be predicted by the (time independent) decay rate from a local friction term. In turn this slow-down of energy transfer and its explicit  time dependence from transient dynamics may affect the balance of energy densities between different species, perhaps impacting the recent results of refs.\cite{dienes} on unexpected dynamical fixed points in   the cosmological evolution of several species as a consequence of a balance between energy transfer  among the different species and the cosmological expansion. These aspects merit further and deeper study.

\vspace{1mm}

\section{ Conclusions:}\label{sec:conclusions}

This study is motivated by the ubiquity and  importance of scalar (or pseudoscalar) condensate decay in post inflationary cosmology, from reheating to axionic dark matter. Our objective is to study the relaxation of such condensate as a consequence of (stimulated) emission and absorption of massless quanta in equilibrium with a thermal bath during the radiation dominated era, within which most of the relevant particle physics processes occur. With this purpose, we introduce a model in which a massive scalar field develops an homogeneous expectation value and interacts with a massless field in thermal equilibrium with the radiation background. The relaxation of the condensate amplitude is a consequence of the stimulated emission and absorption of the massless quanta.

The equation of motion for the condensate is usually appended with a \emph{phenomenological} time independent local friction term, with a friction coefficient obtained from the Minkowski space-time S-matrix decay rate,  to describe the relaxation of the condensate in the expanding cosmology. We argue that an S-matrix calculation, which considers a transition probability per unit time from an in-state prepared at time $t \rightarrow -\infty$ to an out-state at time $t \rightarrow +\infty$, is not a suitable characterization of relaxational processes in an expanding cosmology. In particular, the lack of a global time-like   Killing vector prevents strict energy conservation in the process, yet the S-matrix approach neglects this important aspect of a time dependent cosmology taking infinite time limits thereby enforcing strict energy conservation in the process.

Instead, after a consistent field quantization in the radiation dominated cosmology,  we obtain the causal (retarded) equations of motion including the non-local one-loop self-energy correction from the massless particles in the medium.  The dynamical renormalization group is implemented to obtain the decay function  of the condensate amplitude, directly comparing the result to that obtained from the phenomenological friction term, and also to the Minkowski space-time limit.

The decay law of the condensate is obtained analytically in two regimes: the super-Hubble and sub-Hubble regimes wherein the Compton wavelength of the massive particle is either much larger or much smaller than the Hubble radius respectively, in both cases providing a direct comparison with a local friction term. In the super-Hubble limit, $mt \ll 1$, we find that the condensate amplitude in comoving time decays as $e^{-\frac{g^2}{10} t^2\,\ln(1/mt)}$ where $g$ is proportional to the coupling, showing that the friction term is unsuitable to describe the dynamics of the condensate in this limit. In the sub-Hubble limit, $mt > 1$, the condensate amplitude decays as $e^{-\gamma(t,T(t))}$ where $T(t)=T/a(t)$ is the temperature redshifted by the scale factor, hence at long time the decay is completely dominated by the zero temperature contribution. An important result is that even in the sub-Hubble limit while $\gamma(t)$ approaches asymptotically the zero temperature decay function from the local friction term for $mt \gg 1 $, the phenomenological friction term \emph{does not} reliably describe the decay of the condensate throughout the cosmological evolution, missing important transient phenomena at early and intermediate time scales, and always \emph{underestimating} its lifetime. In particular, for ultralight dark matter decaying into radiation, the time scale during which transient phenomena is sensitive to the cosmological expansion and during which a local friction term is inadequate to describe the decay of the condensate may be very long.

In this study we focused on the simpler case of a homogeneous condensate of a light massive field decaying into massless particles, which in a radiation dominated cosmology are conformally coupled to gravity.   We thoroughly compared the solutions of the correct equations of motion obtained \emph{ab initio} featuring a non-local self-energy kernel to the phenomenological equation with a local friction term,\emph{ calculated in Minkowski space-time}, under the \emph{same} approximations, namely linearized equations of motion, massless decay products and  to leading order in the coupling. Our results clearly indicate that even in this simpler case  the local friction term calculated in Minkowski space-time is inadequate to describe the radiative relaxation of the condensate. Taking the decay products (namely the $\phi_2$ field) to be massive entails  a daunting endeavor even at the linearized level and leading order in the coupling because the one-loop self energy  must be calculated with the Weber functions (\ref{U},\ref{Ucc}). However, it seems unlikely that massive particles in the loop will change the main conclusions on important transient dynamics and consequences of cosmological expansion in particular for ultralight condensates as is the focus of this study. Going beyond the linearized approximation entails even more complexities as the dynamical renormalization group must be extended to include not only expansion in the coupling but also in the amplitude. Therefore, while both extensions, massive fields in the loop and non-linear equations of motion undoubtedly merit further study, a main message of this article is that the simple phenomenological addition of a friction term obtained via S-matrix in Minkowski space-time and  added to the equation of motion to describe radiative relaxation in an expanding cosmology must be skeptically scrutinized, and its applicability to describe the damping of homogeneous condensates from particle decay in cosmology should not be taken for granted.

\vspace{1mm}

\textbf{Further questions:} The results of this study, in particular the consequences of the quantization and the lack of strict energy conservation as a consequence of the cosmological expansion, raise important questions on a Boltzmann equation approach to relaxation and thermalization. This is an integro-differential equation for the time evolution of the distribution function of particles that involves a collision kernel which is typically obtained by calculating transition probabilities per unit time from the S-matrix approach. Such a calculation is based on Minkowski space-time mode functions $e^{\pm i \omega_k t}$ and in taking the infinite time limit which enforces strict energy conservation. As explicitly shown in section (\ref{sec:quantization}) the mode functions are very different from those in Minkowski space-time,  transient dynamics is sensitive to cosmological expansion and the time interval, determined by the Hubble time is finite allowing transient phenomena that does not satisfy strict energy conservation.  As discussed above, the decay of the condensate into radiation is a mechanism of energy transfer between species, therefore the transient dynamics studied in this article may bear impact on the balance between energy transfer and expansion that leads to the emergent dynamical attractor in the study of ref.\cite{dienes}. These subtle aspects merit further study to obtain a consistent framework to assess the time evolution of distribution functions and energy transfer early in a  radiation dominated cosmology.

Our study focused on the time evolution and decay of the homogeneous condensate, namely the expectation value of the scalar or pseudoscalar field motivated by its importance in post-inflationary cosmology, but does not address the question of correlation functions and the approach to thermalization. In reference \cite{shuboy}  non-equilibrium effective action and quantum master equation frameworks were introduced to study the evolution of condensates, correlation functions and thermalization in Minkowski space-time. Extending these methods to a radiation dominated cosmology would offer a pathway towards understanding the dynamics of correlation functions and the approach to thermalization in an expanding cosmology.

\acknowledgements
  The authors thank Keith Dienes and Brooks Thomas for enlightening discussions. They gratefully acknowledge  support from the U.S. National Science Foundation through grants   NSF 2111743 and NSF 2412374.

\appendix
\section{Equation of motion in Minkowski space-time:}\label{app:eomlap}
The Minkowski space-time limit is obtained from the FRW case by taking $a(\eta) = H_R \,\eta \rightarrow 1$ thereby setting the metric to be the Minkowski one, and $\eta \rightarrow t$ in the expressions obtained for the radiation dominated cosmology. Therefore the equation (\ref{fineom}) for the condensate (``zero mode'') in Minkowski space-time becomes

\be \ddot{X}(t)+m^2\,X(t)+  \int^t_0 {\Sigma}(t-t')\,X(t')\,dt' =0 \,,\label{eomXmink}\ee
where
\be \Sigma(t-t') = -2i\lambda^2 \,\int \frac{1}{4k^2}\,[1+2\,n(k)]\,\Big[e^{-2ik(t-t')}-e^{2ik(t-t')}  \Big] \,d^3k \,, \label{sigmink}\ee which we write in a spectral representation as
\be  \Sigma(t-t') = -i \int^{\infty}_{-\infty} \rho(k_0)\,e^{-ik_0(t-t')}\,\frac{dk_0}{(2\pi)} \,,\label{spec}\ee with
\be \rho(k_0) = (4\pi\,\lambda^2) \int \frac{1}{4k^2}\,\Big[1+2\,n(k)\Big]\,\Big[ \delta(k_0-2k)-\delta(k_0+2k)\Big]\,d^3k\,.\label{rhosd} \ee We find
\be \rho(k_0) =  {\pi\,g^2}  \,\Big[1+2\,n\Big(\frac{|k_0|}{2}\Big)\Big] \,\mathrm{sign}(k_0) ~~;~~ g^2 = \frac{\lambda^2}{4\pi^2}\,.\label{rhosi}\ee
The equation of motion (\ref{eomXmink}) can be solved by Laplace transform  as befits an initial value problem. Introducing the Laplace transforms
\bea \widetilde{X}(s) & = &  \int^\infty_0 e^{-st}\,X(t)\,dt \,,\label{laplaX}    \\ \overline{\Sigma}(s) & = &  \int^\infty_0 e^{-st}\,\Sigma(t)\,dt  = -  \,\int^{\infty}_{-\infty} \frac{\rho(k_0)}{k_0-is} \frac{dk_0}{2\pi} \,,\label{laplasigma}
\eea where in (\ref{laplasigma}) we used the dispersive representation (\ref{spec}).
The solution of the Laplace transform   is
 \be \widetilde{X}(s) = \frac{\dot{X}(0) +s\,X(0)}{s^2+m^2+\overline{\Sigma}(s)}\,.\label{laplasolution}\ee
The solution in real time is obtained by inverse Laplace transform, it is given by
\be X(t)   =  X(0)\,\dot{\mathcal{G}}(t) + \dot{X}(0)\,\mathcal{G}(t)   \,,\label{realtisol} \ee  where $\mathcal{G}(t)$ is given by
\be \mathcal{G}(t) = \frac{1}{2\pi i} \int_{\mathcal{C}} \frac{e^{st}}{s^2+m^2+\overline{\Sigma}(s)}\, ds \,, \label{goftsol}\ee   $\mathcal{C}$ denotes the Bromwich contour parallel to the imaginary axis   and to the right of all the singularities   of $(s^2+m^2+\overline{\Sigma}(s))^{-1}$ in the complex s-plane and closing along a large semicircle at infinity with $Re(s)<0$. These singularities correspond to poles and multiparticle branch cuts with $Re(s)<0$, thus the contour runs parallel to the imaginary axis $s= i(\nu -i \epsilon)$, with $-\infty \leq \nu \leq \infty$ and $\epsilon \rightarrow 0^+$. Therefore,
\be \mathcal{G}(t) = - \int^{\infty}_{-\infty} \widetilde{\mathcal{G}}(\nu)\, {e^{i\nu\,t}} \,\frac{d\nu}{2\pi}\,, \label{Goftfin}\ee  where
\be \widetilde{\mathcal{G}}(\nu) = \frac{1}{(\nu-i\epsilon)^2 -m^2 - \overline{\Sigma}(\nu) }\,, \label{Gfnu}  \ee is recognized as the retarded propagator at zero spatial momentum.
The self energy in frequency space is given by the dispersive form
\be \Sigma (\nu)   =    \,\int^{\infty}_{-\infty}  \,  \frac{\rho(k_0)}{\nu-k_0-i\epsilon} \,\frac{dk_0}{2\pi} \equiv \Sigma_R(\nu) + i \Sigma_I(\nu) \,,
 \label{signu}\ee with the real and imaginary parts given by
 \bea \Sigma_R(\nu) & = &   \,\mathcal{P}\int^{\infty}_{-\infty}   \Bigg[ \frac{\rho(k_0)}{\nu-k_0}\Bigg]\,\frac{dk_0}{2\pi} \,,\label{resig}\\
  \Sigma_I(\nu) & = & \frac{1}{2}\,\rho(\nu) \,, \label{imsig} \eea

  To obtain the above representations we have used the relation $\rho(-k_0) = -\rho(k_0)$ (see eqn. (\ref{rhosi})), as a consequence of which it follows that $\Sigma_R(\nu) = \Sigma_R(-\nu)~;~\Sigma_I(\nu) = - \Sigma_I(-\nu)$. $\widetilde{\mathcal{G}}(\nu)$ given by eqn. (\ref{Gfnu})   features   complex poles corresponding to the solution of the equation
  \be \nu^2_P = m^2  + \overline{\Sigma}(\nu_P)\,, \label{poleG}\ee to leading order in $g^2$ we find
  \be \nu_P = \pm m_R+ i \frac{\Gamma}{2}\,, \label{polval}\ee where
  \be m_R = m + \frac{\overline{\Sigma}_R(m)}{2m} ~~;~~ \Gamma = \frac{\Sigma_I(m)}{m}=  \,\frac{\rho(m)}{2m} =\frac{ {\pi\,g^2}}{2m}  \,\Big[1+2\,n\Big(\frac{m}{2}\Big)\Big] \,. \label{reimpole}\ee

  Writing in the denominator of the integrand in (\ref{Goftfin}) $\overline{\Sigma}(\nu) = \overline{\Sigma}(\nu_p) + (\overline{\Sigma}(\nu)-\overline{\Sigma}(\nu_p))$ we find that near each pole, $\widetilde{\mathcal{G}}_k(\nu)$ can be written in
  a Breit-Wigner form as
  \be \widetilde{\mathcal{G}}(\nu) =  \frac{Z}{2\nu_P(\nu \mp m_R -i\frac{\Gamma}{2})} \,\,, \label{BW} \ee with the wave function renormalization constant
   \be Z^{-1} = 1- \frac{\overline{\Sigma}'(m_R)}{2m_R} = 1 + \mathcal{O}(g^2)~,~~~ \overline{\Sigma}'(m_R) \equiv \Big[\frac{d}{d\nu}\overline{\Sigma}(\nu)\Big]_{\nu=m_R}\,. \label{zwf} \ee
    To leading order in $g^2$ we find
   \be \mathcal{G}(t) = e^{-\frac{\Gamma}{2}t}\,\frac{\sin(m_R t)}{m_R } + \mathcal{O}(g^2)\,, \label{goftbw} \ee where we have assumed a narrow width $\Gamma/m_R \propto g^2 \ll 1$ and neglected terms of this order.

   Writing the initial values in terms of complex amplitudes, we find

   \be X(t) = X(0)\, e^{-im_R t}\, e^{-\frac{\Gamma}{2}t} + c.c  \label{propasol1} \,,\ee where $\Gamma$ is given by eqn. (\ref{reimpole}) and the renormalization of the mass features a logarithmic dependence on an ultraviolet cutoff from the zero temperature contribution to the spectral density as can be explicitly gleaned from
   the expression (\ref{resig}).

   To establish a direct relation with the dynamical renormalization group approach discussed in the next subsection, let us consider solving for $\mathcal{G}(t)$ (\ref{goftsol}) in strict perturbation theory in $\overline{\Sigma} \propto g^2$,
   \be \mathcal{G}(t) = \frac{1}{2\pi i} \int_{\mathcal{C}}  {e^{st}}\Bigg\{\frac{1}{s^2+m^2} -\frac{ \overline{\Sigma}(s)}{(s^2+m^2)^2}+ \frac{\overline{\Sigma}^2(s)}{(s^2+m^2)^3}+ \cdots\Bigg\}\, ds \ee the $n^{th}$ power  of $\overline{\Sigma}$ feature denominators with poles or order $n+1$ at $s = \pm i m$, at long time the contributions from the multiparticle cuts of $\overline{\Sigma}$ vanish by dephasing (Riemann-Lebesgue) and the poles dominate,  yielding
   \be  \mathcal{G}(t) = \frac{e^{imt}}{2i}\Big[1 -it\,\overline{\Sigma}(m)+ \frac{(-it)^2}{2}\, \overline{\Sigma}^2(m)+\cdots \Big] + c.c. \ee These terms that grow in time are secular terms, the Dyson geometric sum of the Green's function $\mathcal{G}(t)$ (\ref{goftsol}) provides a \emph{resummation} of these secular terms. The dynamical renormalization group, discussed in detail in the next subsection performs this resummation directly in real time, thereby bypassing the Dyson geometric series which can only be
   performed in Fourier or Laplace space, which are not available in an expanding cosmology.

\section{Dynamical renormalization group in Minkowski space-time}\label{app:drg}
 Let us re-write the equation of motion in a manner so that we can implement perturbation theory and the dynamical renormalization group
 following the method of reference\cite{drg}

\be \ddot{X}(t)+m^2\,X(t)+g^2 \int^t_0 \widetilde{\Sigma}(t-t')\,X(t')\,dt' =0 \,,\label{eomXmink2}\ee where $\Sigma(t-t') \equiv g^2 \widetilde{\Sigma}(t-t')$.

Here, we do not explicitly include a mass counterterm to show how mass renormalization emerges naturally from the dynamical renormalization group.

In perturbation theory, the solution to the equation of motion (\ref{eomXmink2}) is
\be X(t)= X_0(t) + g^2 X_1(t) + \cdots \,,\label{ptsol}\ee yielding the hierarchy of equations
\bea \ddot{X}_0(t) + m^2\,X_0(t) & = & 0 \,,\label{zerod}\\
\ddot{X}_1(t) + m^2\,X_1(t) & = & - \int^t_0 \widetilde{\Sigma}(t-t')\,X_0(t')\,dt' \,,\label{firsd}\\
 \vdots ~~~~~~ & = & ~~~~~~\vdots \eea
The solution to the zeroth order equation (\ref{zerod}) is
\be X_0(t) = X\,e^{-imt}+X^*\,e^{imt} \,,\label{X0sol} \ee where $X$ is a complex amplitude, and the solution to the first order equation (\ref{firsd}) is
\be X_1(t) = -\int^\infty_0  G_R(t-t')\int^{t'}_0 \widetilde{\Sigma}(t'-t^{''})\,X_0(t^{''}) dt^{''}\, dt' \,,\label{X1sol}\ee where the retarded Green's function
\be G_R(t-t') = \frac{1}{m}\,\sin[m(t-t')]\,\Theta(t-t') \,,\label{Gretmink}\ee

\subsection{Zero temperature}\label{app:subTzero}

The zero temperature self-energy in Minkowski space-time is simply given by eqn. (\ref{I0}) with the replacement $\eta \rightarrow t$, and absorbing the factor $1/8\pi$ into the definition of the coupling $g$, namely
\be \widetilde{\Sigma}(t-t') = -  \frac{t-t'}{(t-t')^2+\epsilon^2}~~;~~ \epsilon \rightarrow 0\,,\label{SEMink}\ee   Let us first calculate the integral on the right hand side of eqn. (\ref{firsd}),
\be \int^t_0 \widetilde{\Sigma}(t-t')\,X_0(t')\,dt' = - X\,e^{-imt} \, \mathcal{F}(t)-X^*\, e^{imt}\,\mathcal{F}^*(t) \,,\label{intfirst} \ee where the zero temperature contribution is
\be \mathcal{F}(t) = \int^{t}_0 \frac{\tau}{\tau^2+\epsilon^2}\,e^{im\tau} \,d\tau\,,\label{calF}\ee  which we write in terms of the variable $z \equiv m\tau$ as
\be \mathcal{F}(t) = \int^{mt}_0 \frac{z}{z^2+\epsilon^2m^2}\,dz + \int^{mt}_0   \frac{\cos(z)-1}{z}\,dz + i \int^{mt}_0   \frac{\sin(z)}{z} \,dz\,,\label{foftex}\ee where in the second and third terms we set $\epsilon =0$ since there is no singularity as $z \rightarrow 0$, yielding for $\epsilon \rightarrow 0^+$ and finite $t$
\be \mathcal{F}(t) =-\ln(\epsilon\,m)+\ln(mt) +  \int^{mt}_0   \frac{\cos(z)-1}{z}\,dz + i \int^{mt}_0   \frac{\sin(z)}{z} \,dz\,.\label{foftex2}\ee

Introducing the ultraviolet cutoff $\Lambda = 1/\epsilon$, $\mathcal{F}(t)$ becomes
\be \mathcal{F}(t) =  \ln\Big( \frac{\Lambda}{m}\Big)-\gamma + Ci[mt] + i \, Si[mt] \,,\label{finft} \ee where $\gamma=0.577\cdots$ is   Euler's constant, and $Ci,Si$ are the cosine and sine integral functions respectively, which feature the long time limit
\be Ci[mt]~~{}_{\overrightarrow{mt \gg 1 }} ~~ 0 ~~;~~ Si[mt]~~{}_{\overrightarrow{mt \gg 1 }} ~~ \frac{\pi}{2} \,,\label{asyfot}\ee therefore, in the long time limit
\be \mathcal{F}(t) ~~{}_{\overrightarrow{mt \gg 1 }} ~~  \ln\Big( \frac{\Lambda}{m}\Big)-\gamma + i\,\frac{\pi}{2} \,.\label{flontime}\ee

Inserting the result (\ref{intfirst}) into equation (\ref{X1sol}) we find
\bea X_1(t)  & = & X\,e^{-imt}\,\Big[1+ i \frac{g^2}{2m}\,\int^t_0 \mathcal{F}(t')\,dt' \Big] + X^*\,e^{imt}\,\Big[1- i \frac{g^2}{2m}\,\int^t_0 \mathcal{F}^*(t')\,dt' \Big] \nonumber \\
& - & i\,\frac{g^2}{2m} \, X \,e^{imt}\,  \int^t_0  e^{-2imt'} \mathcal{F}(t')\,dt'   + i\,\frac{g^2}{2m} X^*\,e^{-imt}\, \int^t_0 e^{2imt'} \mathcal{F}(t')\,dt' \,.\label{X1fini}\eea In the long time limit the first line features a secular term that grows linearly in time from the integrals, whereas the contributions from the second line are not secular in the long time limit. The brackets in the first line suggest a renormalization of the amplitudes, hence, following refs.\cite{drg1,drg} we write
\be X \equiv \widetilde{X}(\overline{t})\,Z(\overline{t})~~;~~ Z[\overline{t}] = 1 + g^2\,z_1(\overline{t}) + \cdots \label{XofT}\ee where $\overline{t}$ is a renormalization scale, therefore
\be X(t) = \widetilde{X}(\overline{t}) \,e^{-imt}\,\Big[1+ i \frac{g^2}{2m}\,\int^t_0 \mathcal{F}(t')\,dt' + g^2\,z_1(\overline{t}) \Big] + c.c.  + \mathrm{non~secular} \,,\label{Xren}\ee choosing
\be z_1(\overline{t})= -\frac{i}{2m} \int^{\overline{t}}_0 \mathcal{F}(t')\,dt'\,,\label{z1choice}\ee it follows that
\be X(t) = \widetilde{X}(\overline{t}) \,e^{-imt}\,\Big[1+ i \frac{g^2}{2m}\,\int^{t}_{\overline{t}} \mathcal{F}(t')\,dt' \Big] + c.c + \mathrm{non~secular} \,,\label{XrenT} \ee
 therefore the perturbative solution has been improved by choosing $\overline{t}$ to be near $t$. Since the scale $\overline{t}$ is arbitrary and the solution $X(t)$ is independent of this scale, it  obeys the dynamical renormalization group equation
\be \frac{d X({t})}{d\overline{t}} = 0 \,,\label{drgeqmink} \ee
 which up to $\mathcal{O}(g^2)$ yields
\be \frac{d \widetilde{X}(\overline{t})}{d\overline{t}} = \frac{i g^2}{2m} \mathcal{F}(\overline{t})\,\widetilde{X}(\overline{t}) \,,\label{drgX}\ee with solution
\be \widetilde{X}(\overline{t}) = \widetilde{X}(\overline{t}_0)\,e^{ \frac{i g^2}{2m}  \int^{\overline{t}}_{\overline{t}_{0}}\mathcal{F}(\overline{t}')\,d\overline{t}'}\,. \label{soluXfina}\ee
Choosing $\overline{t}\equiv t~;~\overline{t}_0=0$ and inserting the solution (\ref{drgX}) into (\ref{XrenT}) we finally find in the long time limit
\be X(t) = X(0)\, e^{-im_R t}\, e^{-\frac{\Gamma}{2}t} + c.c + (\mathrm{perturbative}) \,,\ee where the renormalized mass $m_R$ and decay width $\Gamma$ are given by
\be m_R = m -  \frac{g^2}{2m}\Big[ \ln\Big(\frac{\Lambda}{m}\Big) -\gamma \Big]~~;~~ \Gamma = \frac{\pi\,g^2}{2m} \,.\label{masswidth}\ee

This analysis explicitly shows the power of the dynamical renormalization group, it directly yields the time evolution of the amplitude and correctly describes the asymptotic long time limit in direct agreement with the S-matrix result which is obtained from the exact Dyson resummed propagator, equation (\ref{propasol1}).

\subsection{Finite temperature in real time}\label{app:subTnonzero}
From the results (\ref{IT}) the finite temperature contribution to the self energy $\widetilde{\Sigma}$  in Minkowski space time is given by
\be \widetilde{\Sigma}_T(t-t') = -2 \sum_{l=1}^{\infty} \frac{(t-t')}{\Big[ \big( \frac{l}{2T}\big)^2+ (t-t')^2 \Big]} \,,\label{sigT}\ee
Using the identity (\cite{grad}),
\be \coth\Big[\pi y \Big] = \frac{1}{\pi y}+ \frac{2}{\pi}\,\sum_{l=1}^{\infty}\, \frac{y}{y^2+ l^2}\,\label{ident}\ee
we find
\be \widetilde{\Sigma}_T(t-t') = - (2\pi T) \Big[\coth\big[2\pi T (t-t')\big]   - \frac{1}{2\pi T (t-t')}\Big]\,.\label{sigTcot} \ee The finite temperature contribution to $\mathcal{F}_T(t)$ in eqn. (\ref{intfirst}) is given by
\be \mathcal{F}_T(t) = 2 \,\sum_{l=1}^{\infty}\,\int^{t}_0  \frac{\tau}{\Big[ \big( \frac{l}{2T}\big)^2+ \tau^2 \Big]}\,e^{im\tau} \,d\tau\,,\label{calFT}\ee  which in terms of the variable $z=mt$ becomes
\be \mathcal{F}_T(t) = 2 \,\sum_{l=1}^{\infty}\,\int^{mt}_0  \frac{z}{\Big[ \big( \frac{l\,m}{2T}\big)^2+ z^2 \Big]}\, \Big( \cos(z) + i \sin(z) \Big)  \,dz\,,\label{calFTz}\ee
We are primarily interested in the long time limit $mt \rightarrow \infty$ of the sine term which yields the decay rate, for which we use
\be \int^\infty_{0} \frac{z}{z^2+a^2} \sin(z) dz = \frac{1}{2i} \int^{\infty}_{-\infty} \frac{z\,e^{iz}}{z^2+ a^2}\,  dz = \frac{\pi}{2}\,e^{-a}~~;~~a>0\,,\label{intelt}\ee  yielding
\be \mathrm{Im}[\mathcal{F}_T] = \frac{\pi}{2} \Bigg[ 2 \,\sum_{l=1}^{\infty} e^{-lm/2T}\Bigg] = \frac{\pi}{2} \Big[ 2\,n\Big(\frac{m}{2}\Big) \Big]~~;~~ n(x) = \frac{1}{e^{\frac{x}{T}}-1} \,,\label{gamaT} \ee which when combined with the zero temperature contribution (\ref{masswidth}) yields for the decay rate
\be \Gamma = \frac{\pi\,g^2}{2m} \Bigg[1+  2\,n\Big(\frac{m}{2}\Big)\Bigg]\,,\label{gamatot}\ee which coincides with the more familiar solution of the equations of motion in terms of Fourier-Laplace transforms as in the previous subsection, where the decay rate emerges from the imaginary part of the self-energy at the position of the pole in the Green's function, see equation (\ref{reimpole}).

\acknowledgements
  The authors  gratefully acknowledge  support from the U.S. National Science Foundation through grants   NSF 2111743 and NSF 2412374.

\end{document}